\RequirePackage{rotating}
\documentclass[acmsmall]{acmart}
\usepackage{graphicx}
\usepackage{subfig}
\usepackage{amsmath}
\usepackage{gensymb}
\usepackage{makecell}
\usepackage{tabularx}
\usepackage{multirow}
\usepackage{lipsum}
\usepackage{enumitem}
\usepackage{xcolor}
\usepackage{colortbl}
\usepackage[symbol]{footmisc}

\usepackage{amsmath,amssymb}

\setcellgapes{4pt}

\AtBeginDocument{%
  \providecommand\BibTeX{{%
    \normalfont B\kern-0.5em{\scshape i\kern-0.25em b}\kern-0.8em\TeX}}}

\setcopyright{acmcopyright}
\copyrightyear{2018}
\acmYear{2018}
\acmDOI{10.1145/1122445.1122456}

\acmJournal{JACM}
\acmVolume{37}
\acmNumber{4}
\acmArticle{111}
\acmMonth{8}

\begin{document}

\title{Deep Learning for Sensor-based Human Activity Recognition: Overview, Challenges and Opportunities}

\author{Kaixuan Chen}
\authornote{Both authors contributed equally to the paper}
\affiliation{
  \institution{Aalborg University}
  \city{Aalborg}
  \postcode{9220}
  \country{Denmark}}
\email{kchen@cs.aau.dk}

\author{Dalin Zhang}
\authornotemark[1]
\orcid{0000-0002-5869-6544}
\affiliation{%
  \institution{Aalborg University}
  \city{Aalborg}
  \postcode{9220}
  \country{Denmark}}
\email{dalinz@cs.aau.dk}

\author{Lina Yao}
\affiliation{%
  \institution{University of New South Wales}
  \city{Sydney}
  \state{NSW}
  \postcode{2052}
  \country{Australia}}
\email{lina.yao@unsw.edu.au}

\author{Bin Guo}
\affiliation{%
  \institution{Northwestern Polytechnical University}
  \city{Xi'an}
  \state{Shaanxi}
  \postcode{710129}
  \country{China}}
\email{guobin.keio@gmail.com}

\author{Zhiwen Yu}
\affiliation{%
  \institution{Northwestern Polytechnical University}
  \city{Xi'an}
  \state{Shaanxi}
  \postcode{710129}
  \country{China}}
\email{zhiwenyu@nwpu.edu.cn}

\author{Yunhao Liu}
\affiliation{%
  \institution{Michigan State University}
  \city{East Lansing}
  \state{MI}
  \postcode{48824}
  \country{USA}}
\email{yunhao@cse.msu.edu}

\renewcommand{\shortauthors}{K. Chen et al.}

\begin{abstract}
The vast proliferation of sensor devices and Internet of Things enables the applications of sensor-based activity recognition. However, there exist substantial challenges that could influence the performance of the recognition system in practical scenarios. Recently, as deep learning has demonstrated its effectiveness in many areas, plenty of deep methods have been investigated to address the challenges in activity recognition. In this study, we present a survey of the state-of-the-art deep learning methods for sensor-based human activity recognition. We first introduce the multi-modality of the sensory data and provide information for public datasets that can be used for evaluation in different challenge tasks. We then propose a new taxonomy to structure the deep methods by challenges. Challenges and challenge-related deep methods are summarized and analyzed to form an overview of the current research progress. At the end of this work, we discuss the open issues and provide some insights for future directions.
\end{abstract}



\maketitle

\section{Introduction}
Recent advance in human activity recognition has enabled myriad applications such as smart homes \cite{ishimaru2017towards026}, healthcare \cite{li2016deep031}, and enhanced manufacturing \cite{grzeszick2017deep122}.
Activity recognition is essential to humanity since it records people's behaviors with data that allows computing systems to monitor, analyze, and assist their daily life.
There are two mainstreams of human activity recognition systems: video-based systems and sensor-based systems. Video-based systems use cameras to take images or videos to recognize people's behaviors \cite{azar2019convolutional}. Sensor-based systems utilize on-body or ambient sensors to dead reckon people's motion details or log their activity tracks. Considering the privacy issues of installing cameras in our personal space, sensor-based systems have dominated the applications of monitoring our daily activities. Besides, sensors take advantage of pervasiveness. Thanks to the proliferation of smart devices and Internet of Things, sensors can be embedded in portable devices such as phones, watches, and nonportable objects like cars, walls, and furniture.
Sensors are widely embedded around us, uninterruptedly and non-intrusively logging human's motion information.

\subsection{Challenges in Human Activity Recognition.}
\begin{figure}[ht]
\includegraphics[width=\textwidth]{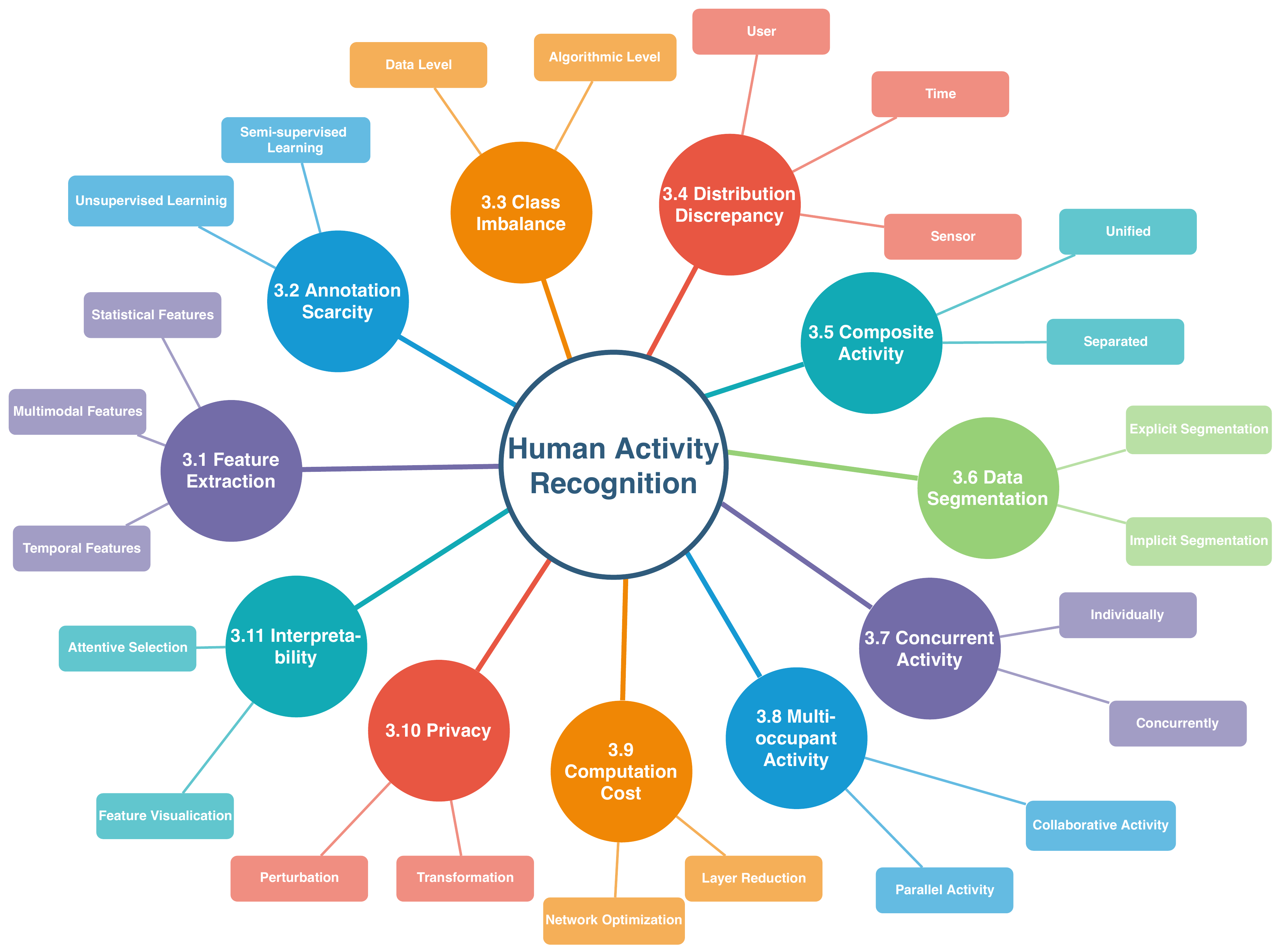}
\caption{Categories of deep learning in sensor based human activity recognition}
\label{fig:taxonomy}
\end{figure}

Many machine learning methods have been employed in human activity recognition. However, this field still faces many technical challenges. Some of the challenges are shared with other pattern recognition fields such as computer vision and natural language processing, while some are unique to sensor-based activity recognition and require dedicated methods for real-life applications. Here lists a few categorizes of challenges that the community of activity recognition should respond. A figure of the taxonomy is shown in Figure ~\ref{fig:taxonomy}.

\begin{itemize}[leftmargin=*]
    \item The first challenge is \textbf{\textit{feature extraction}}. Activity recognition is a classification task so it shares a common challenge with other classification problems which is feature extraction. For sensor-based activity recognition, feature extraction is more difficult because there is inter-activity similarity \cite{bulling2014tutorial004}. Different activities may have similar characteristics (e.g., walking and running). Therefore, it is difficult to produce distinguishable features to represent activities uniquely.
    
    \item Training and evaluation of learning techniques require large annotated data samples. However, it is expensive and time-consuming to collect and annotate sensory activity data. Therefore, \textbf{\textit{annotation scarcity}} is a remarkable challenge for sensor-based activity recognition. Besides, data for some emergent or unexpected activities (e.g., accidentally fall) is especially hard to obtain, which leads to another challenge called \textbf{\textit{class imbalance}}.
    
    \item Human activity recognition involves three factors: users, time, and sensors. First, activity patterns are person-dependent. Different users may have diverse activity styles. Second, activity concepts vary over time. The assumption that users remain their activity patterns unchanged in a long time is impractical. Moreover, novel activities are likely to emerge when in use. Thirdly, diverse sensor devices are opportunistically configured on human bodies or in environments. The composition and the layouts of sensors dramatically influence the data stimulated by activities. All the three factors lead to \textbf{\textit{distribution discrepancy}} between the training data and test data and need to be mitigated urgently.
    
    \item The complexity of data association is another reason that makes recognition challenging. Data association refers to how many users and how many activities the data is associated with. There are many specific challenges in activity recognition that are driven by sophisticated data association.
    The first challenge can be seen in \textit{\textbf{composite activities}}. Most activity recognition tasks are based on simple activities, like walking and sitting. However, more meaningful ways to log human daily routines are composite activities that comprise a sequence of atomic activities. For example, ``washing hands'' can be represented as $\{$turning on the tap, soaping, rubbing hands, turning off the tap$\}$. 
    One challenge stimulated by composite activities is \textit{\textbf{data segmentation}}. A composite activity can be defined as a sequence of activities. Therefore, accurate recognition highly relies on precise data segmentation techniques.
    \textit{\textbf{Concurrent activities}} show the third challenge. Concurrent activities occur when a user participates in more than one activities simultaneously, such as answering a phone call while watching TV.
    \textit{\textbf{ Multi-occupant activities}} are also associated with the complexity of data association. Recognition is arduous when multiple users engage in a set of activities, which usually happens in multi-residents scenarios. 
    
    \item Another factor that needs to be concerned is the feasibility of the human activity recognition system.
    Efforts need to be devoted to making the system acceptable by a vast number of users since human activity recognition is quite close to human daily life, which can be twofold. 
    First, the system should be recourse-intensive so that it fits portable devices and is able to give an instant response. Thus, the \textbf{\textit{computational cost}} issue should be addressed. Second, as the recognition system records users' life continuously, there are risks of personal information disclosure, which leads to the \textbf{\textit{privacy}} issue.
    
    \item Unlike images or texts, sensory data is unreadable. Moreover, sensory data inevitably includes lots of noise information on account of the inherent imperfections of sensors.
    So, reliable recognition solutions should have \textbf{\textit{interpretability}} in sensory data and the capability of understanding which part of data facilitates recognition and which part deteriorates that.
\end{itemize}

\subsection{Deep Learning in Human Activity Recognition.}  
Numerous previous works adopted machine learning methods in human activity recognition \cite{lara2013survey}. They highly rely on feature extraction techniques including time-frequency transformation \cite{huynh2005analyzing}, statistical approaches \cite{bulling2014tutorial004} and symbolic representation \cite{lin2003symbolic}.
However, the features extracted are carefully engineered and heuristic. There were no universal or systematical feature extraction approaches to effectively capture distinguishable features for human activities.

In recent years, deep learning has embraced conspicuous prosperity in modeling high-level abstractions from intricate data \cite{pouyanfar2018survey003} in many areas such as computer vision, natural language processing, and speech processing.
After early works including \cite{ha2015multi110,yang2015deep017,lane2015can114} examined the effectiveness of deep learning in human activity recognition, related studies sprung up in this area.
Along with the inevitable development of deep learning in human activity recognition, latest works are undertaken to address the specific challenges. 
However, deep learning is still confronted with reluctant acceptance by researchers
owing to its abrupt success, bustling innovation, and lack of theoretical support.
Therefore, it is necessary to demonstrate the reasons behind the feasibility and success of deep learning in human activity recognition despite the challenges.

\begin{itemize}[leftmargin=*]
    \item The most attractive characteristic of deep learning is ``deep''. Layer-by-layer structures of deep models allow to learn from simple to abstract features scalably. Also, advanced computing resources like GPUs provide deep models with a powerful ability to learn descriptive features from complex data. The outstanding learning ability also enables the activity recognition system to analyze multimodal sensory data for accurate recognition deeply.
    
    \item Diverse structures of deep neural networks encode features from multiple perspectives. For example, convolutional neural networks (CNNs) are competent in capturing the local connections of multimodal sensory data, and the translational invariance introduced by locality leads to accurate recognition \cite{hammerla2016deep016}.
    Recurrent neural networks (RNNs) extract the temporal dependencies and incrementally learn information through time intervals so are appropriate for streaming sensory data in human activity recognition.
    
    \item Deep neural networks are detachable and can be flexibly composed into unified networks with one overall optimization function, which makes allowance for miscellaneous deep learning techniques including deep transfer learning \cite{akbari2019transferring_024}, deep active learning \cite{gudur2019activeharnet081}, deep attention mechanism \cite{murahari2018attention029} and other not systematic but as effective solutions \cite{mathur2018using_023,ito2018application043}. Works that adopted these techniques cater to various challenges in deep learning.
\end{itemize}

\subsection{Key Contributions.}
Unlike the existing surveys related to deep learning in human activity recognition, we focus distinctly on 
the \textit{challenges} of human activity recognition and how motivated deep learning models and techniques are developed to be challenge-specific. Specifically, Wang et al. \cite{wang2019deep002} surveyed a number of deep learning methods for sensor-based human activity recognition in the view of model structures. Nweke et al. \cite{nweke2018deep072} presented a survey only on mobile and wearable sensor-based activity recognition and categorized the deep learning methods into generative, discriminative, and hybrid methods. Li et.al \cite{li2019survey126} introduced different deep neural networks for radar-based activity recognition. These surveys only discuss the deep models that can be used for activity recognition (e.g. CNNs and RNNs) while we expand the scope to the techniques that can be well merged with deep learning to tackle specific challenges (e.g. deep transfer learning, multimodal fusion).

Compared with the existing surveys, the key contributions of this work can be summarized as follows:

\begin{itemize}[leftmargin=*]
    \item We conduct a comprehensive survey of deep learning approaches for sensor-based human activity recognition. Our work provides a panorama of current progress and an in-depth analysis of the reviewed methods to serve both novices and experienced researchers.
    \item We propose a new taxonomy of deep learning methods in the view of challenges of activity recognition. Challenges stimulated by different reasons are presented for the readers to scan which research direction is of interest. 
    \item We summarize the state-of-the-art and how specific deep networks or deep techniques can be applied to address the challenges with comprehensive analysis. We compare different solutions for the same challenges and list the pros and cons. The challenge-method-analysis format aims to build a problem-solution structure with a hope to suggest a rough guideline when readers are selecting their research topics or developing their approaches. 
    \item Moreover, we provide information on available public datasets and their potential extension to evaluate specific challenges. 
    \item We discuss some open issues in this field and point out potential future research directions.
\end{itemize}

\section{Sensor Modality and Datasets}

\subsection{Sensor Modality}
The performance of an activity recognition system depends crucially on the used sensor modality. In this section, we classify the sensor modalities into four strategies: wearable sensors, ambient sensors, object sensors, and other modalities. 

\subsubsection{Wearable Sensor}
As wearable sensors can directly and efficiently capture body movements, they are the most commonly used for human activity recognition. These sensors can be freely integrated into smartphones, watches, bands, and even clothes.

\noindent\textbf{\textit{Accelerometer.}} An accelerometer is a device used to measure acceleration which is the rate of change of the velocity of an object. The measuring unit is meters per second squared ($m/s^2$) or G-forces ($g$). The sampling frequency is usually in the range of tens to hundreds of Hz. For recognizing human activities, accelerometers can be mounted on various parts of a body, such as the waist \cite{anguita2013public_UCIHAR}, arm \cite{zappi2008activity_Skoda}, ankle \cite{bachlin2010wearable_Daphnet}, wrist \cite{huynh2008discovery_Darmstadt}, et al. There are three axes in an often-used accelerometer. Therefore, a tri-variate time series would be achieved through an accelerometer.

\noindent\textbf{\textit{Gyroscope.}} A gyroscope is a device that measures orientation and angular velocity. The unit of angular velocity is measured in degrees per second (\degree$/s$). The sampling rate is also from tens to hundreds of Hz. A gyroscope is usually integrated with an accelerometer and amounted on the same body parts. In addition, a gyroscope has three axes
as well.

\noindent\textbf{\textit{Magnetometer.}} A magnetometer is another widely used wearable sensor for activity recognition, which is generally assembled with an accelerometer and a gyroscope into an inertial unit. It measures the change of a magnetic field at a particular location. The measurement units are Tesla ($T$), and the sampling rate is from tens to hundreds of Hz. Likewise, a magnetometer has three axes.

\noindent\textbf{\textit{Electromyography (EMG).}} An EMG sensor is used to evaluate and record the electrical activity produced by skeletal muscles. Different from the above three kinds of sensors, EMG sensors require to be attached directly to human skin. As a result, it is less commonly used in conventional scenarios but more suitable for fine-grained motions such as hand \cite{zia2018multiday} or arm \cite{wu2015real} movements and facial expressions. The EMG provides a univariate time series of signal amplitudes.

\noindent\textbf{\textit{Electrocardiography (ECG).}} ECG is another biometric tool for activity recognition that measures the electrical activities generated by the heart. It also requires the sensor to contact the human's skin directly. As different people's hearts vibrate in significantly different ways, the ECG signals are difficult for processing subject variations. An ECG sensor provides a univariate time series data. 

\subsubsection{Ambient Sensor}
Ambient sensors are usually embedded in the environment to capture the interactions between humans and the environment. A unique advantage of ambient sensors is that they can be used to detect multi-occupant activities. In addition, the ambient sensors can also be adopted for in-door localizing, which is difficult for wearable sensors to achieve.

\noindent\textbf{\textit{WiFi.}} WiFi is a local-area wireless network connection technology which uses a transmitter to send signals to a receiver. The basis of the WiFi-based human activity recognition is that human's movements and locations interfere with the signals' propagation path from the transmitter to the receiver, including both the direct propagation path and the reflecting propagation path. 

\noindent\textbf{\textit{Radio-frequency identification (RFID).}} RFID uses electromagnetic fields to automatically identify and track the tags attached to objects, which contains electronically stored information. There are two kinds of RFID tags: active and passive tags. Active tags rely on a local power source (such as a battery) to continuously broadcast their signals that can be detected hundreds of meters away from an RFID reader. In contrast, passive RFID tags collect energy from a nearby RFID reader's interrogating radio waves to send its stored information. 
Thus, passive RFID tags are much cheaper and lighter. RSS is the mostly adopted tool for RFID-based activity recognition \cite{yao2017compressive,wang2018modeling}. The working mechanism is that human's movements would change the single strength received by the RFID reader.

\noindent\textbf{\textit{Radar.}} Different from WiFi and RFID whose transmitters and receivers are placed on the opposite sides, radar transmitters and antennas are mounted on the same side of users. Doppler effect is the basis of the radar-based system \cite{li2019survey126}. 

\subsubsection{Object Sensor}
The wearable and ambient sensors are used to target the motions of humans themselves. However, besides simple activities (e.g., walking, sitting, jogging et al.), human performs composite activities (e.g., drinking/eating, cooking, playing et al.) through continuously interacting with surroundings in practical scenarios. As a result, incorporating the information on using objects is crucial for recognizing more complex human activities.

\noindent\textbf{\textit{Radio-frequency identification (RFID).}} Regarding the cost-efficiency, reliability, and easy implementation, RFID sensors are the most widely used for identifying object usage. When acting as object sensors rather than ambient sensors, RFID tags are needed to be attached to the target objects such as mugs, books, computers, and toothpaste \cite{buettner2009recognizing}. In the detection phase, a worn RFID reader is also needed. 
The reading of an object sensor is processed to be binary marks for indicating whether the object is used.

\subsubsection{Other Modalities} In addition to the above sensor modalities, there are other modalities that have particular applications.

\noindent\textbf{\textit{Audio Sensor.}} Modern mobile devices normally have a built-in pair of a speaker and a microphone, which can be used to recognize human activities. The speaker is used to transmit ultrasound signals, and the microphone is used to receive the ultrasound signals. The basis is that the ultrasound would be modified by human movements and thus reflects the motion information. 
This modality is particularly suitable for recognizing human's fine-grained movements as control commands of mobile devices since no external devices or signals are required \cite{ruan2017making}.

\noindent\textbf{\textit{Pressure Sensor.}} Unlike the above ambient sensing modalities which use electromagnetic or sound waves to grasp human activities, the pressure sensor depends on mechanical mechanisms, which requires direct physical contact. It can be embedded in either smart environments or wearable equipment. When implanted in the smart environment, pressure sensors can be deposited at diverse places, such as a chair \cite{cheng2016smart}, a table \cite{cheng2016smart}, a bed \cite{foubert2012lying}, and the floor \cite{rangarajan2007design}. Due to its characteristics of physical contact, small movements or various static postures can be detected. 
Therefore, it may be suitable for particular scenarios like exercise monitoring
\cite{cheng2016smart} and writing posture corrections \cite{lee2018analysis}.

\subsection{Datasets}
There are several publicly available human activity recognition datasets.
We summarize some of the most popular ones in Table \ref{tab:dataset}, which contains the data acquisition context, number of subjects, number of activities, sensor types, and potential challenge tasks they can be used in. In the data acquisition context, "daily living" refers to subjects performing common daily living activities under instructions. The challenges are further detailedly explained in Section 3.

\begin{sidewaystable}
\caption{Public Datasets for Human Activity Recognition}
\footnotesize
\centering
\begin{tabular}{cccccc}
    \toprule
     Dataset & Context & \# Subject & \# Activities & Sensor Types & Challenges\\\toprule
     WISDM Activity Prediction \cite{kwapisz2011activity_WISDM} & Daily Living & 29 & 6 & Wearable & Class Imbalance \\
     UCI HAR \cite{anguita2013public_UCIHAR} & Daily Living & 30 & 6 & Wearable & Multimodal \\
     \multirow{3}{*}{OPPORTUNITY \cite{chavarriaga2013opportunity,roggen2010collecting_Opportunity}} & \multirow{3}{*}{Daily Living} & \multirow{3}{*}{4} & \multirow{3}{*}{9} & \multirow{3}{*}{Wearable, Object, Ambient} & \multirow{3}{*}{\makecell{Multimodal\\ Composite Activity}}\\\\\\\\
     Skoda Checkpoint \cite{zappi2008activity_Skoda}& Car Maintenance & 1 & 10 & Wearable & Simple\\
     Daphnet Freezing of Gait \cite{bachlin2010wearable_Daphnet} & Patients of Parkinson's Disease & 10 & 3 & Wearable & Simple \\
     Berkeley MHAD & Daily Living & 12 & 11 & Wearable, Ambient & Multimodal \\
     PAMAP2 \cite{reiss2012introducing_PAMAP2} & Daily Living & 9 & 18 & Wearable & Multimodal\\
     SHO \cite{shoaib2014fusion_SHO} & Daily Living & 10 & 7 & Wearable & Simple \\
     UCI HAPT \cite{reyes2016transition_UCIHAPT} & Daily Living with activity transition & 30 & 6 & Wearable & Multimodal \\
     UTD-MHAD \cite{chen2015utd_UTD-MHAD} & Controlled Conditions & 8 & 27 & Wearable & Multimodal\\
     HHAR \cite{stisen2015smart_HHAR} & Daily Living & 9 & 6 & Wearable & Multimodal, Distribution Discrepancy \\
     ARAS \cite{alemdar2013aras_ARAS} & Real-world Home Living & 2 & 27 & Ambient, Object & Multimodal, Multi-occupant\\
     Ambient Kitchen \cite{pham2009slice_AmbientKitchen} & Food Preparation & 20 & 11 & Object & Simple \\
     USC-HAD \cite{zhang2012usc_USC-HAD} & Daily Living & 14 & 12 & Wearable & Multimodal \\
     MHEALTH \cite{banos2014mhealthdroid_MHealth} & Real-world Home Living & 10 & 12 & Wearable & Multimodal \\
     BIDMC Congestive Heart Failure \cite{baim1986survival} & Hear failure & 15 & 2 & Wearable & Class Imbalance \\
     DSADS \cite{barshan2014recognizing_DSADS} & Daily Living and Sports & 8 & 19 & Wearable & Multimodal \\
     \multirow{4}{*}{CASAS-4 \cite{singla2010recognizing}} & \multirow{4}{*}{Real-world Home Living} & \multirow{4}{*}{2} & \multirow{4}{*}{15} & \multirow{4}{*}{Object, Ambient} &\multirow{4}{*}{ \makecell{Multi-occupant\\ Composite Activity\\ Multimodal}}\\\\\\\\
     Smartwatch/Notch/Farseeing \cite{mauldin2018smartfall_Notch} & Daily Living \& Fall Detection & 7 & 4 ADL \& 4 Fall & Wearable & Class Imbalance \\
     Darmstadt Daily Routines \cite{huynh2008discovery_Darmstadt} & Real-world Routines & 1 & 35 & Wearable & Class Imbalance \\
     MotionSense \cite{malekzadeh2019mobile} & Daily Living & 24 & 6 & Wearable & Simple \\
     MobiAct/MobiFall \cite{vavoulas2016mobiact} & Daily Living \& Fall Detection & 66 & 12 ADL \& 4 Fall & Wearable & Multimodal \\
     VanKasteren benchmark \cite{van2011human}& Real-world Home Living & 3 & 9 & Object & Simple\\
     ActiveMiles \footnote{http://hamlyn.doc.ic.ac.uk/activemiles/datasets.html} & Real-world Routines & 10 & 7 & Wearable & Multimodal\\
     ActRecTut \cite{bulling14_csur} & Hand Gesture \& Playing Tennis & 2 & 12 & Wearable & Multimodal \\
\bottomrule    
\end{tabular}
\label{tab:dataset}
\end{sidewaystable}

\section{Challenges and Techniques}

\subsection{Feature Extraction}
\label{feature extraction}
While progress has been made, human activity recognition remains a challenging task. This is partly due to the broad range of human activities and the rich variation in how a given activity can be performed. Using features that clearly separate activities is crucial. Feature extraction is one of the key steps in activity recognition since it can capture relevant information to differentiate various activities. The accuracy of activity recognition approaches dramatically depends on the features extracted from raw signals.
Supervised, semi-supervised, and unsupervised approaches all contribute substantially to human activity recognition. 
After supervised learning proved to be effective in extracting features from activity data \cite{jiang2015human121,ishimaru2017towards026}, a wealth of works on supervised learning have been proposed considering that supervised approaches are more prone to end-to-end training.
To be more organised, in this survey we focus only on supervised learning methods in case of feature extraction.
Unsupervised and semi-supervised learning methods are mainly introduced in case of annotation scarcity.
We summarize feature extraction methods for activity recognition into temporal features, multimodal features, and statistical features.

\subsubsection{Temporal Feature Extraction}
Typically, human activity is a combination of several continuous basic movements and can last from a few seconds to up to several minutes. Therefore, considering the relatively high sensing frequency (tens to hundreds Hz), the data of human activity is represented by time-series signals. In this context, the basic streaming movements are more likely to exhibit a smooth fluctuation, while, in contrast, the transitions between consecutive basic movements may induce substantial changes. In order to capture such signal characteristics of human activities, it is essential to extract temporal features of both within and between successive basic movements.

\begin{figure}[ht]
\setlength {\belowcaptionskip} {-0.2cm}
\subfloat[Time-Frequency-Spectral]{
    \includegraphics[width=.4\textwidth]{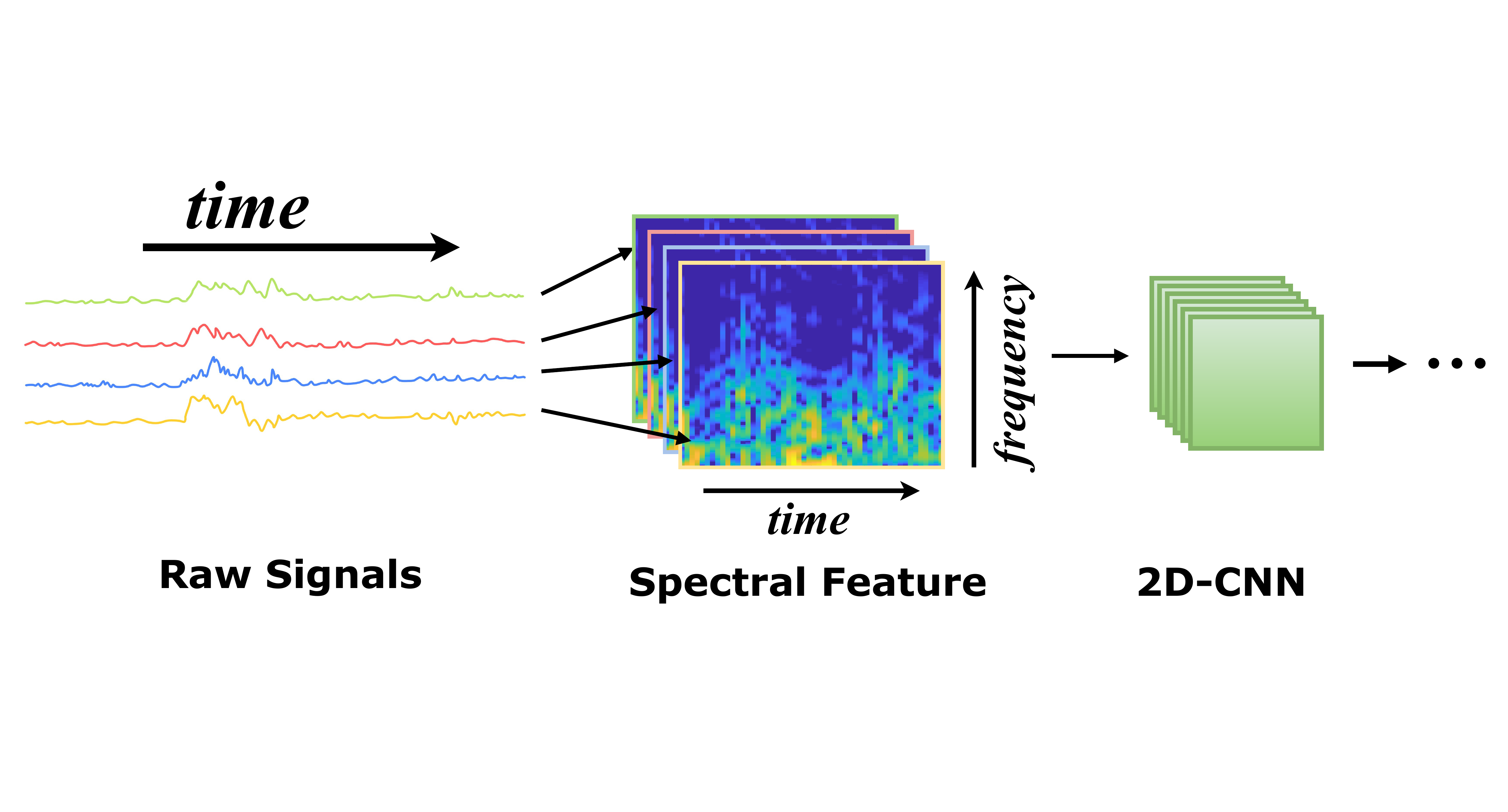}
    }
\subfloat[RNN]{
    \includegraphics[width=.3\textwidth]{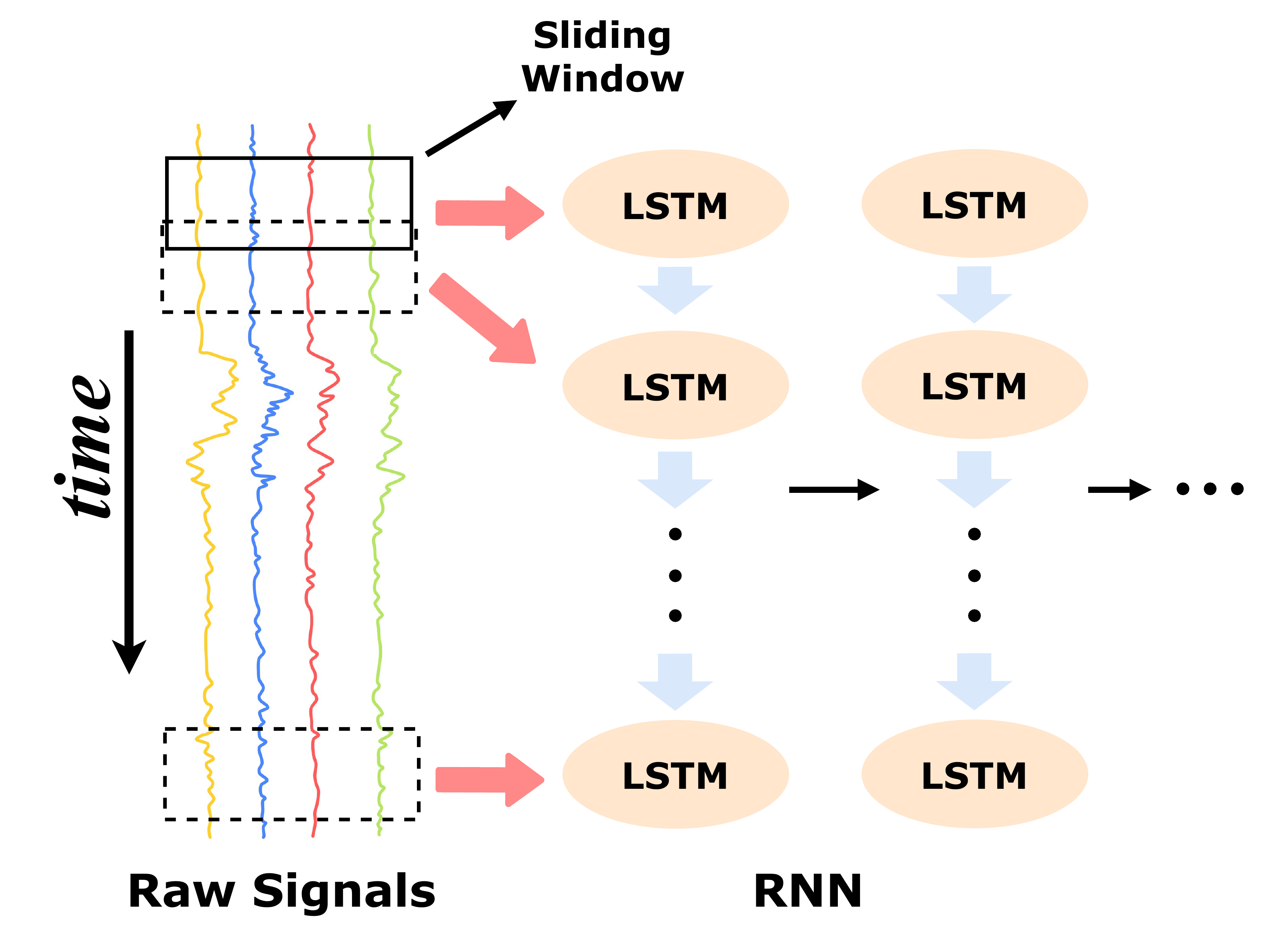}
    }
\subfloat[CNN]{
    \includegraphics[width=.3\textwidth]{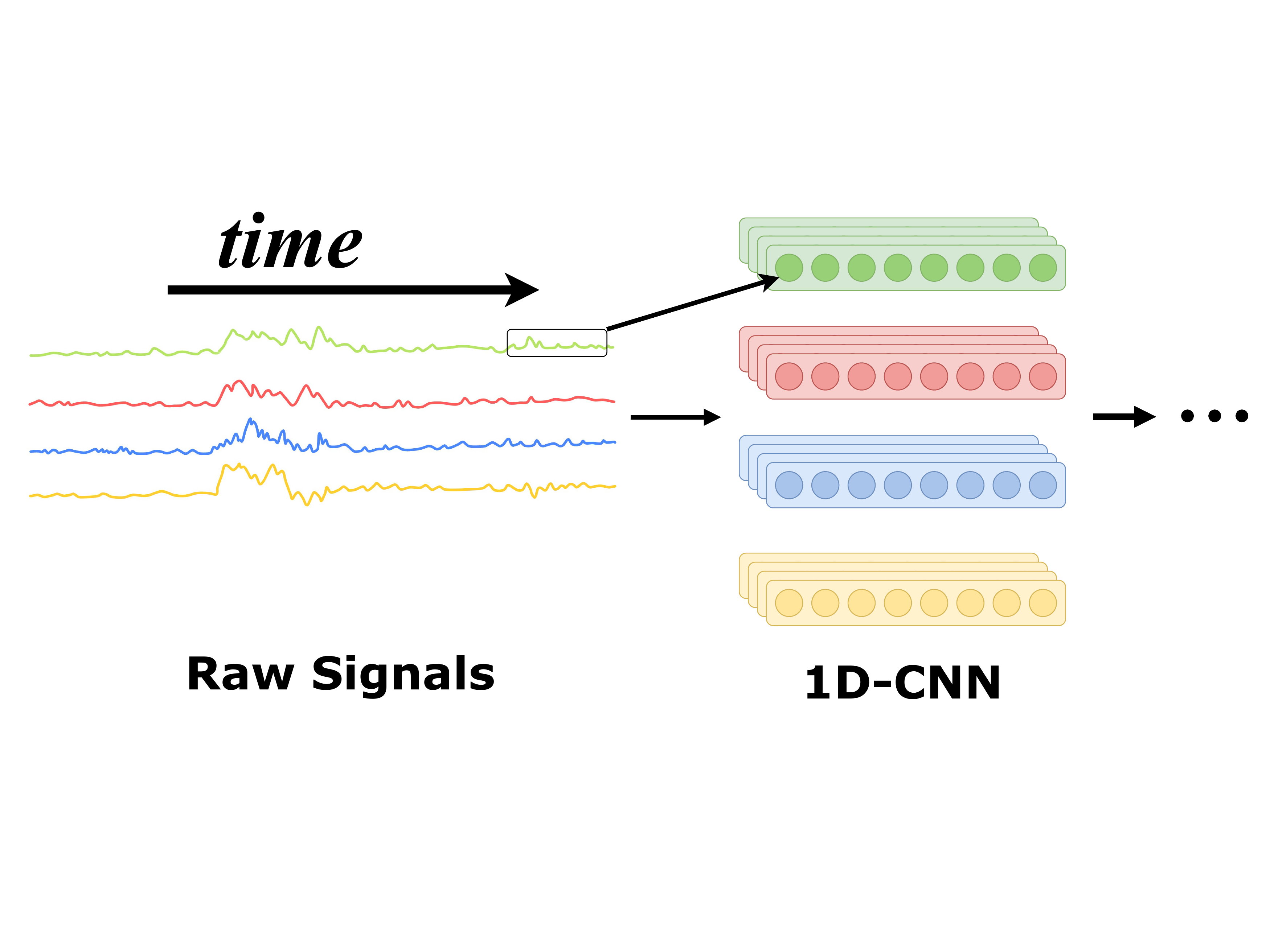}
    }   
\caption{Example structures for temporal feature extraction}
\label{fig:temporal}
\end{figure}

Some researchers manage to adopt traditional methods to extract temporal features and use deep learning techniques for the following activity recognition. Basic signal statistics and waveform traits such as \textit{mean} and \textit{variance} of time-series signals are commonly applied handcrafted features for early-stage deep learning activity recognition \cite{vepakomma2015wristocracy}. This kind of feature is coarse and lacks scalability. A more advanced temporal feature extraction approach is to exploit the spectral power changes as time evolves by converting the time series from the time domain to the frequency domain. A general example structure is shown in Figure~\ref{fig:temporal} (a), where a 2D-CNN is usually used to process the spectral features. In \cite{jiang2015human121}, Jiang and Yin applied the Short-time Discrete Fourier Transform (STDFT) to time-serial signals and constructed a time-frequency-spectral image. Then, CNN is utilized to handle the image for recognizing simple daily activities like walking and standing. More recently, Laput and Harrison \cite{laput2019sensing018} developed a fine-grained hand activity sensing system through the combination of the time-frequency-spectral features and CNNs. They demonstrated 95.2\% classification accuracy over 25 atomic hand activities of 12 people. The spectral features can not only be used for the wearable sensor activity recognition but also be used for the device-free activity recognition. Fan et al. \cite{fan2018tagfree054} proposed to develop time-angle spectrum frames for representing the spectral power variations along time in different spatial angles of the RFID signals.

Since one of the most favorable advantages of the deep learning technology is the impressive power of automatic feature learning, extracting temporal features by a neural network is favorable to construct an end-to-end deep learning model. The end-to-end learning manner facilitates the training procedure and mutually promotes the feature learning and recognition processes. Various deep learning approaches have been applied for temporal information extraction, including RNN, temporal CNN, and their variants. RNN is a widely applied deep temporal feature extraction approach in many fields \cite{zhang2019making,mikolov2010recurrent}. Traditional RNN cells suffer from vanishing/exploding gradients problems, which limits the application of EEG analysis. The Long Short-Term Memory (LSTM) units that have overcome this issue are usually used to build an RNN for temporal feature extraction \cite{greff2016lstm}. The depth of an effective LSTM-based RNN needs to be at least two when processing sequential data \cite{karpathy2015visualizing}. As the sensor signals are continuous streaming data, a sliding window is generally used to segment the raw data to individual pieces, each of which is the input of an RNN cell \cite{chen2016lstm}. A typical LSTM-based structure for temporal feature extraction is illustrated in Figure~\ref{fig:temporal} (b). The length and moving step of the sliding window are hyper-parameters that need to be carefully tuned for achieving satisfying performance. Besides the early application of the basic LSTM network, continuing research of diverse RNN variants is also being investigated in the human activity recognition field. The Bidirectional LSTM (Bi-LSTM) structure that has two conventional LSTM layers for extracting temporal dynamics from both forward and backward directions is an important variant of the RNN in various domains including human activity recognition \cite{ishimaru2017towards026}. In addition, Guan and Pl{\"o}tz \cite{guan2017ensembles051} proposed an ensemble approach of multiple deep LSTM networks and demonstrated superior performance to individual networks on three benchmark datasets. Aside from the variants of the RNN structure, some researchers also studied different RNN cells. For example, Yao et al. \cite{yao2017deepsense047} leveraged the Gated Recurrent Units (GRUs) instead of LSTM cells to construct an RNN and applied it to activity recognition. However, some studies revealed that the other sorts of RNN cells could not provide notably superior performance to the conventional LSTM cell concerning classification accuracy \cite{greff2016lstm}. On the other hand, due to its computational efficiency, GRUs are more suitable for mobile devices where the computation resources are limited.

CNN is another favorable deep learning architecture for temporal feature extraction. Unlike RNN, a temporal CNN does not need a sliding window for segmenting streaming data. The convolution operations with small kernels are directly applied along the temporal dimension of sensor signals so that local temporal dependencies can be captured. Some works employed one-dimensional (1D) convolutions on the individual univariate time series signals for temporal feature extraction \cite{yang2015deep017,duffner20143d119,grzeszick2017deep122,ronao2015deep057,ronao2016human073,bai2020prototype}. When there were multiple sensors or multiple axes, multivariate time series would be yielded, thus requiring the 1D convolutions to be applied separately. Figure~\ref{fig:temporal} (c) presents a typical 1D-CNN structure for temporal feature handling. Conventional 1D CNNs usually have a fixed kernel size, and thus can only discover the signal fluctuations within a fixed temporal range. Considering this gap, Lee et al. \cite{lee2017human115} combined multiple CNN structures of different kernel sizes to obtain the temporal features from different time scales. However, the multi-kernel CNN structure would consume more computational resources, and the temporal scale that a pure CNN could explore is inadequate as well. Furthermore, if a large time scale is desirable, a pooling operation would be commonly used between two CNN layers, which would cause information loss. 
Xi et al. \cite{xi2018deep061} applied a deep dilated CNN to time series for solving the issues. The dilated CNN uses dilated convolution kernels instead of the standard convolutional kernels to expand the convolution receptive field (i.e., time length) with no loss of resolution. Because the dilated kernel only adds empty elements between the elements of the conventional convolution kernel, it does not require an extra computational cost. In addition to the consideration of various temporal scales, the temporal disparity of different sensing modalities (e.g., different sensors, axes, or channels) is also a critical concern since commonly used CNN treats different modalities in the same way. To resolve this concern, Ha and Choi \cite{ha2016convolutional058} presented a new CNN structure that had specific 1D CNNs for different modalities for learning modality-specific temporal characteristics. With the development of the CNNs, other kinds of CNN variants are also considered for effectively embedding temporal features. Shen et al. \cite{shen2018sam093} utilized the gated CNN for daily activity recognition from audio signals and showed superior accuracy to the naive CNN. Long et al. adopted residual blocks to build a two-stream CNN structure dealing with different time scales.

Developing a deep hybrid model to explore different views of temporal dynamics is another attractive trend in the human activity recognition community. In light of the advantages of CNN and RNN, Ord{\'o}{\~n}ez and Roggen \cite{ordonez2016deep111} proposed to combine CNNs and LSTMs for both local and global temporal feature extraction. 
Wang et al. \cite{wang2020push} developed a classifier with a CNN and an LSTM to automatically
extract complicated features from the acoustic data and perform gesture recognition.
Xu et al. \cite{xu2019innohar071} adopted the advanced Inception CNN structure for different scales of local temporal feature extraction and took the GRUs for efficient global temporal representations. Yuki et al. \cite{yuki2018activity038} employed a dual-stream ConvLSTM network with one stream handling smaller time length and the other one handling more substantial time length to analyze more complex temporal hierarchies. Zou et al. \cite{zou2018deepsense} induced an Autoencoder to first enhance feature extractions and then applied the cascade CNN-LSTM to extract local and global features for WiFi-based activity recognition. On the other hand, Gumaei et al. \cite{gumaei2019hybrid} proposed a hybrid model of different types of recurrent units (SRUs and GRUs) for handling different aspects of temporal information.

\begin{figure}[ht]
\setlength {\belowcaptionskip} {-0.5cm}
\subfloat[Feature Fusion]{
    \includegraphics[width=.5\textwidth]{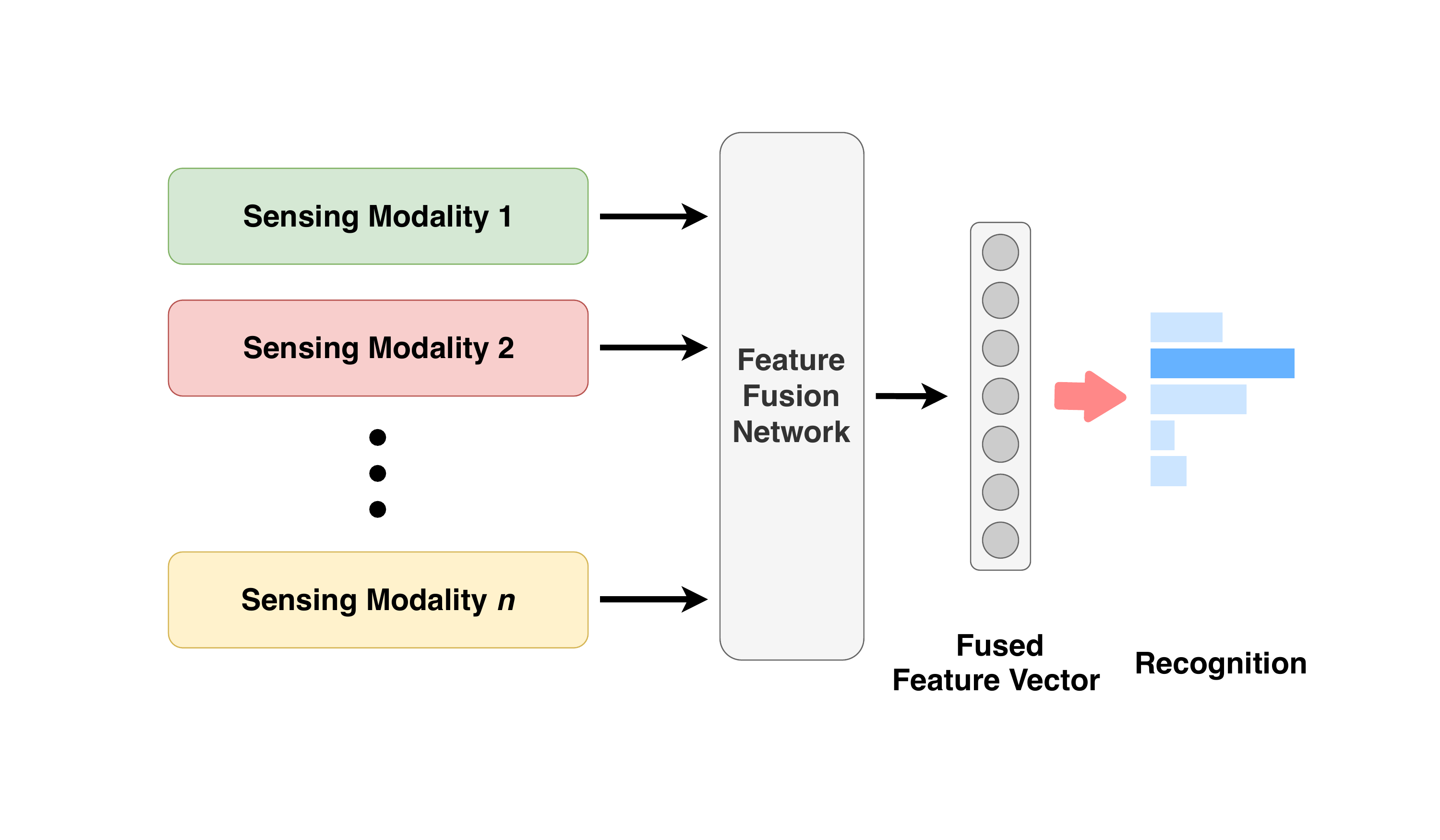}
    }
\subfloat[Classifier Ensemble]{
    \includegraphics[width=.5\textwidth]{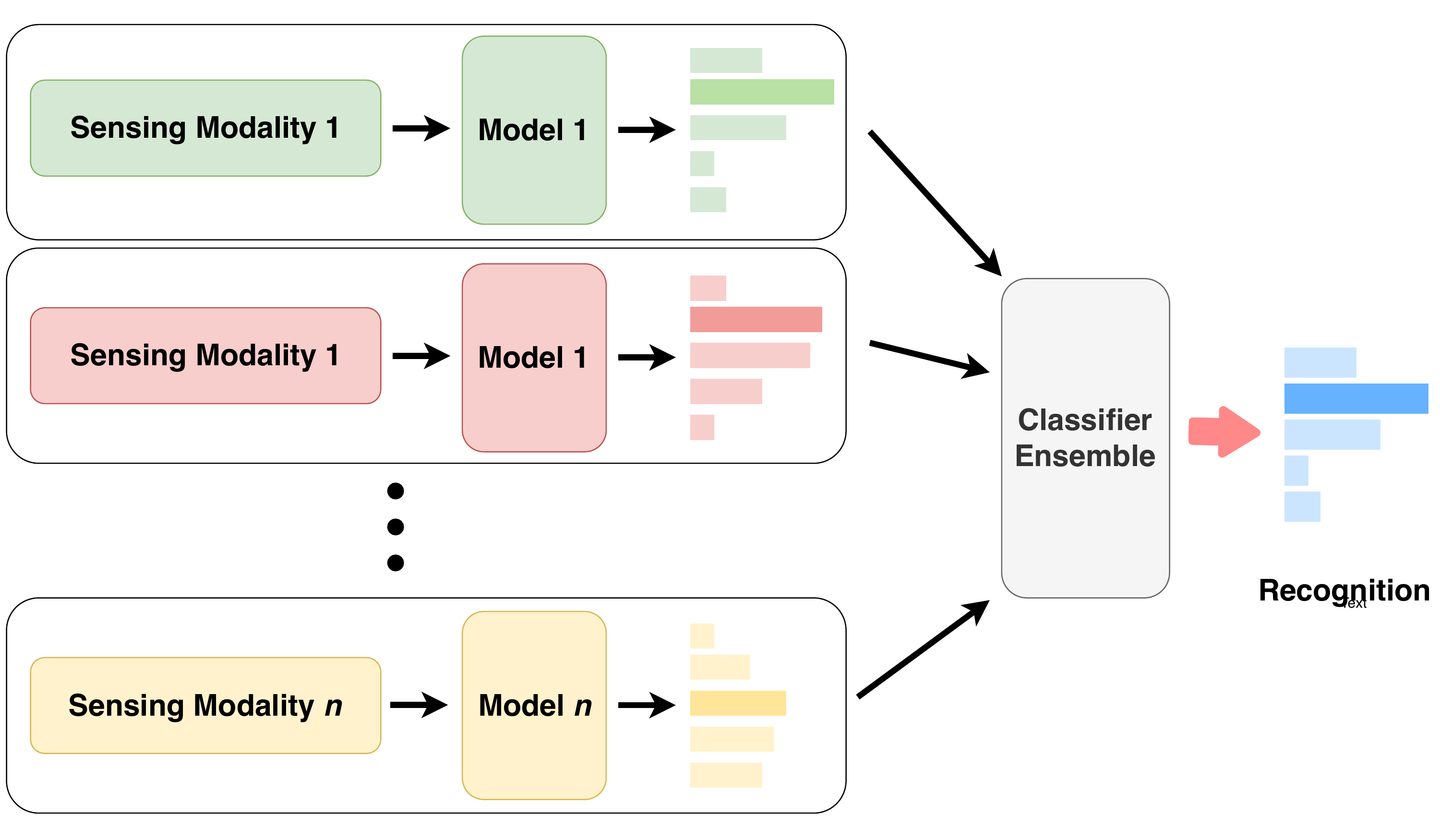}
    }
\caption{Multi-modality fusion strategies}
\label{fig:multimodal}
\end{figure}

\subsubsection{Multimodal Feature Extraction}
The current research of human activity recognition is usually achieved with multiple different sensors, such as accelerometers, gyroscopes, and magnetometers. Some research has further demonstrated that the combination of diverse sensing modalities can obtain better results than one particular sensor only \cite{guo2016wearable046}. As a result, learning the inter-modality correlations along with the intra-modality information is a major challenge in the field of deep learning-based human activity recognition. The sensing modality fusion can be performed following two strategies: \textbf{Feature Fusion} (Figure~\ref{fig:multimodal} (a)) that combines different modalities to produce single feature vectors for classification; and \textbf{Classifier Ensemble} (Figure~\ref{fig:multimodal} (b)) in which outputs of classifiers operating only on features of one modality are blended together. 

M{\"u}nzner et al. \cite{munzner2017cnn028} investigated the feature fusion manner of deep neural networks for multimodal activity recognition. They organized the fusion manners into four categories according to different fusion stages within a network. However, their study focuses on CNN-based architectures only. Here, we extend their definitions of feature fusion manners to all deep learning architectures and manage to reveal more insights and specific considerations.

\textbf{\textit{Early Fusion (EF).}}
This manner fuses the data of all sources at the beginning, irrespective of sensing modalities. It is attractive in terms of simplicity as a strategy though it is at risk of missing detailed correlations. A simple fusion approach in \cite{lee2017human115} transformed the raw $x$, $y$, and $z$ acceleration data into a magnitude vector by calculating the Euclidean norm of $x$, $y$, and $z$ values. Gu et al. \cite{gu2018locomotion075} stacked the time serial signals of different modalities horizontally into a single 1D vector and utilized a denoising autoencoder to learn robust representations. The output of the intermediate layer was used to feed the final softmax classifier. In contrast, Ha et al. \cite{ha2015multi110} proposed to vertically stack all signal sequences to form a 2D matrix and directly applied 2D-CNNs to simultaneously capture both local dependencies over time as well as spatial dependencies over modalities. In \cite{ha2017activity102}, the authors preprocessed the raw signal sequence of a single modality into a 2D format but by simply reorganizing, and stacked all modalities along the depth dimension to finally achieve 3D data matrices. Afterwards, they applied a 3D-CNN to exploit the inter- and intra-modality features. However, conventional CNN is restricted to explore the correlations of neighboring arranged modalities and thus misses the relations between the nonadjacent modalities. To solve this issue, unlike naturally organizing various data sources, Jiang and Yin \cite{jiang2015human121} assembled signal sequences of different modalities into a novel arrangement where every signal sequence has the chance to be adjacent to every other sequence. This organization facilitates the DCNN to extract elaborated correlations of individual sensing axes. Dilated convolution is another solution to exploiting nonadjacent modalities without information loss and extra computational expenses \cite{xi2018deep069}. In addition to wearable sensors, RFID-based activity recognition requires the fusion of multiple RFID signals as well, and CNNs are also commonly used for the early fusion manner \cite{li2016deep037}. 

\textit{\textbf{Sensor-based Fusion (SF).}}
In contrast to EF, SF first considers each modality individually and then fuses different modalities afterwards. Such an architecture not only extracts modality-specific information from various sensors but also allows flexible complexity distribution since the structures of the modality-specific branches can be different. In \cite{radu2018multimodal049,radu2016towards045}, Radu et al. proposed a fully-connected deep neural network (DNN) architecture to facilitate the intra-modality learning. Independent DNN branches are assigned to each sensor modality, and a unifying cross-sensor layer merges all the branches to uncover the inter-modality information. Yao et al. \cite{yao2017deepsense047} vertically stacked all axes of a sensor to form 2D matrices and designed individual CNNs for each 2D matrix to learn the intra-modality relations. The sensor-specific features of different sensors are then flattened and stacked into a new 2D matrix before being fed into a merge CNN for further extracting the interactions among different sensors. A more advanced fusion approach was proposed by Choi et al. \cite{choi2018confidence041} to efficiently fuse different modalities by regulating the level of contribution of each sensor. The authors designed a confidence calculation layer for automatically determining the confidence score of a sensing modality, and then the confidence score was normalized and multiplied with pre-processed features for the following feature fusion of addition. Instead of fusing sensor-specific feature only at the late stage, Ha and Choi \cite{ha2016convolutional058} proposed to create a vector of different modalities at the early stage as well and to extract the common characteristics across modalities along with the sensor-specific characteristics; then both kinds of features are fused at the later part of the model. 

\textit{\textbf{Axis-based Fusion (AF).}} This manner treats signal sources in more detail by handling each sensor axis separately. In such a way, the interference between different sensor axes is gotten rid of. \cite{munzner2017cnn028} referred this manner to \textit{Channel-based late fusion (CB-LF)}. Nevertheless, the sensor channel may be confused with the "channel" in CNNs, so we use the term "axis" instead in this paper. A commonly used AF strategy is to design a specific neural network for each univariate time series of each sensing channel \cite{zheng2014time117,zeng2014convolutional}. The information representations from all channels are concatenated at last for input into a final classification network. 1D-CNNs are widely used as the feature learning network of each sensing channel. Dong and Han \cite{dong2018har094} proposed to use separable convolution operations to extract the specific temporal features of each axis and concatenate all the features before feeding a fully-connected layer. In the studies of applying deep learning to hand-crafted features, the axis-specific process is a requirement. For instance, in \cite{ito2018application043}, temporal features of acceleration and gyro signals are first represented by FFT spectrogram images and then vertically combined into a larger image for the following DCNN to learn inter-modality features. Furthermore, some research combined the spectrogram images along the depth dimension to establish a 3D format \cite{laput2019sensing018}, which could be easily handled by 2D CNNs with the depth dimension as the CNN input channel. 

\textit{\textbf{Shared-filter Fusion (SFF).}} Same to the AF approach, this manner processes the univariate time-serial data of a sensor axis independently. However, the same filter is applied to all time sequences. Therefore, the filters are influenced by all input members. Compared to the AF manner, SFF is more simple and contains fewer trainable parameters. The most popular approach of SFF is to organize the raw sensing sequences into a 2D matrix by stacking along the modality dimension, and then to apply a 2D-CNN to the 2D matrix with 1D filters \cite{yang2015deep017,duffner20143d119,zebin2016human106}. As a result, the architecture is equivalent to applying identical 1D-CNNs to different univariate time series. Although the features of all sensing modalities are not merged explicitly, they communicate with each other by the shared 1D filters.

\begin{figure}[ht!]
\setlength {\belowcaptionskip} {-0.5cm}
\subfloat[Early Fusion]{
    \includegraphics[width=.135\textwidth]{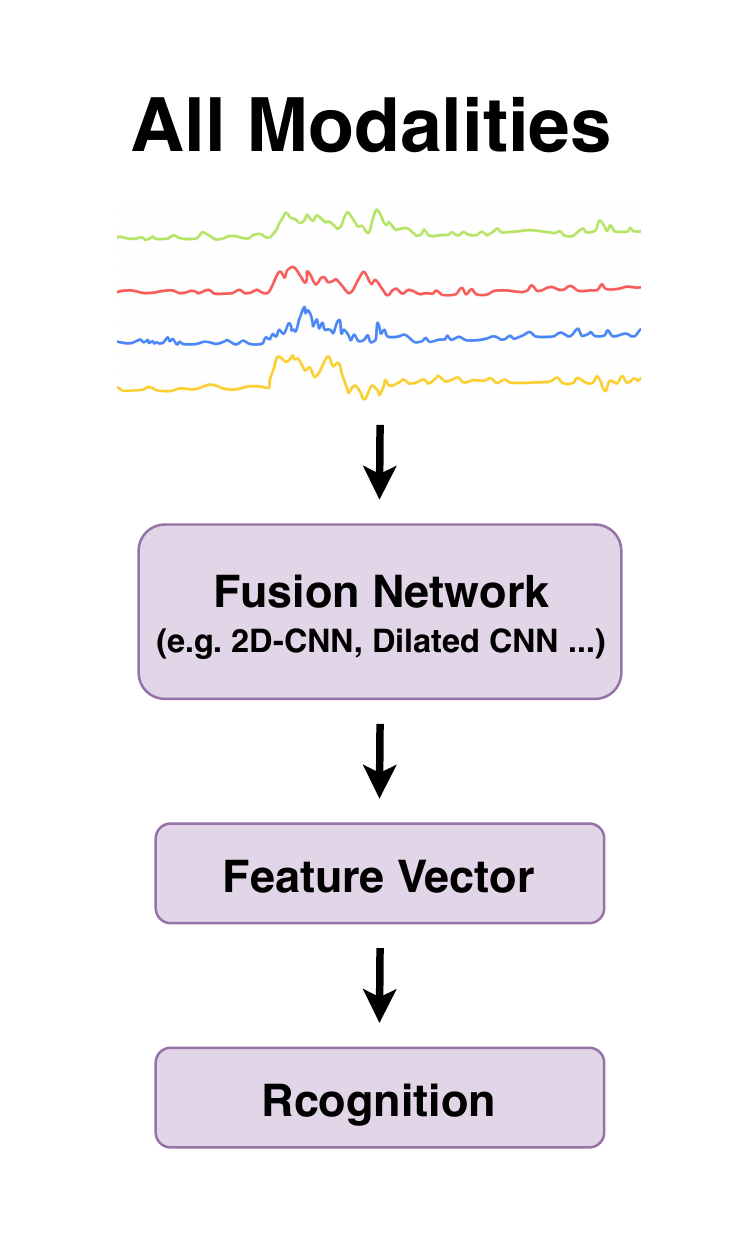}
    \label{fusion-early}}
\subfloat[Sensor-based Fusion]{
    \includegraphics[width=.27\textwidth]{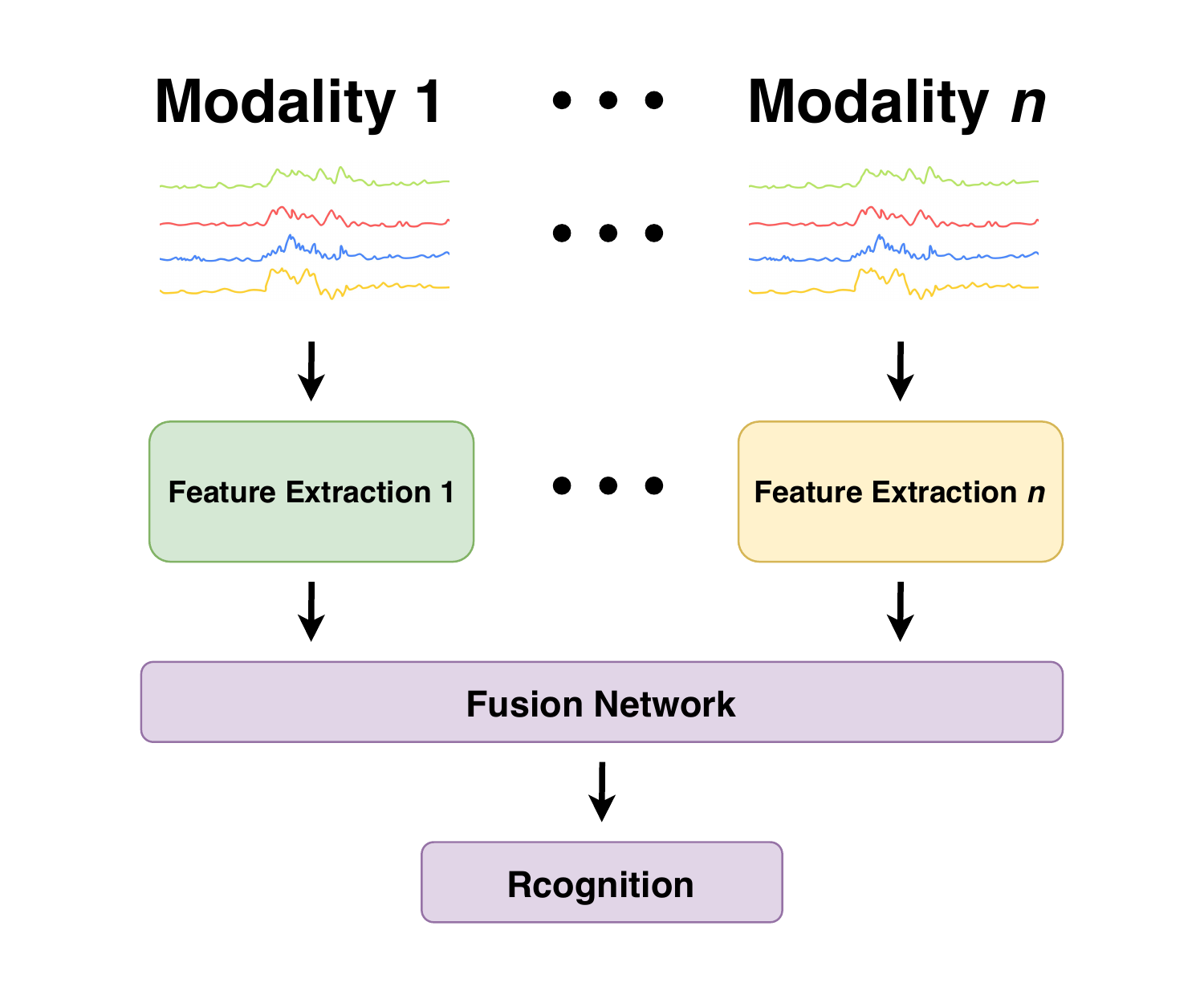}
    \label{fusion-sensor}}
\subfloat[Axis-based Fusion]{
    \includegraphics[width=.27\textwidth]{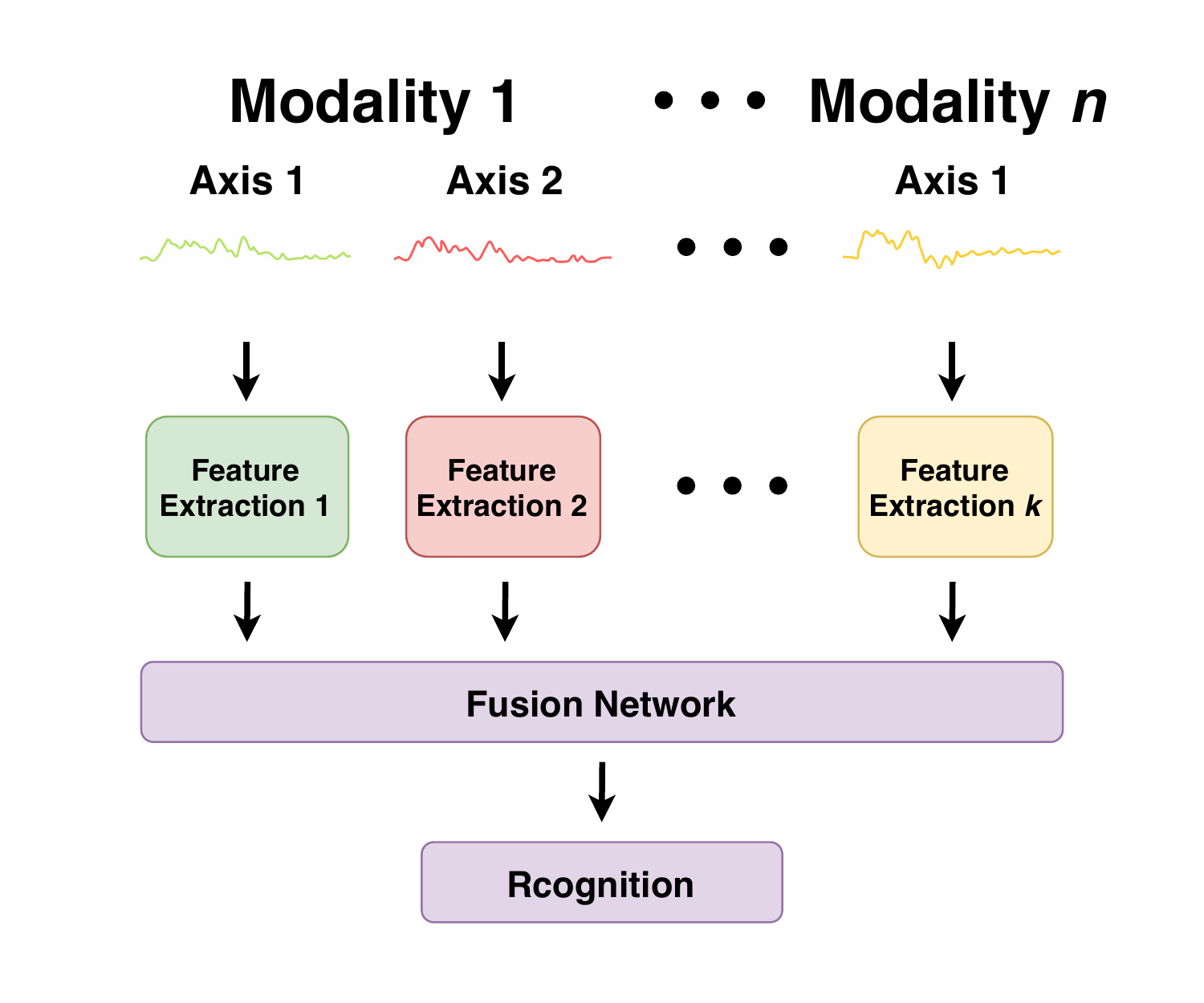}
    \label{fusion_axis}}
\subfloat[Shared-filter Fusion]{
    \includegraphics[width=.27\textwidth]{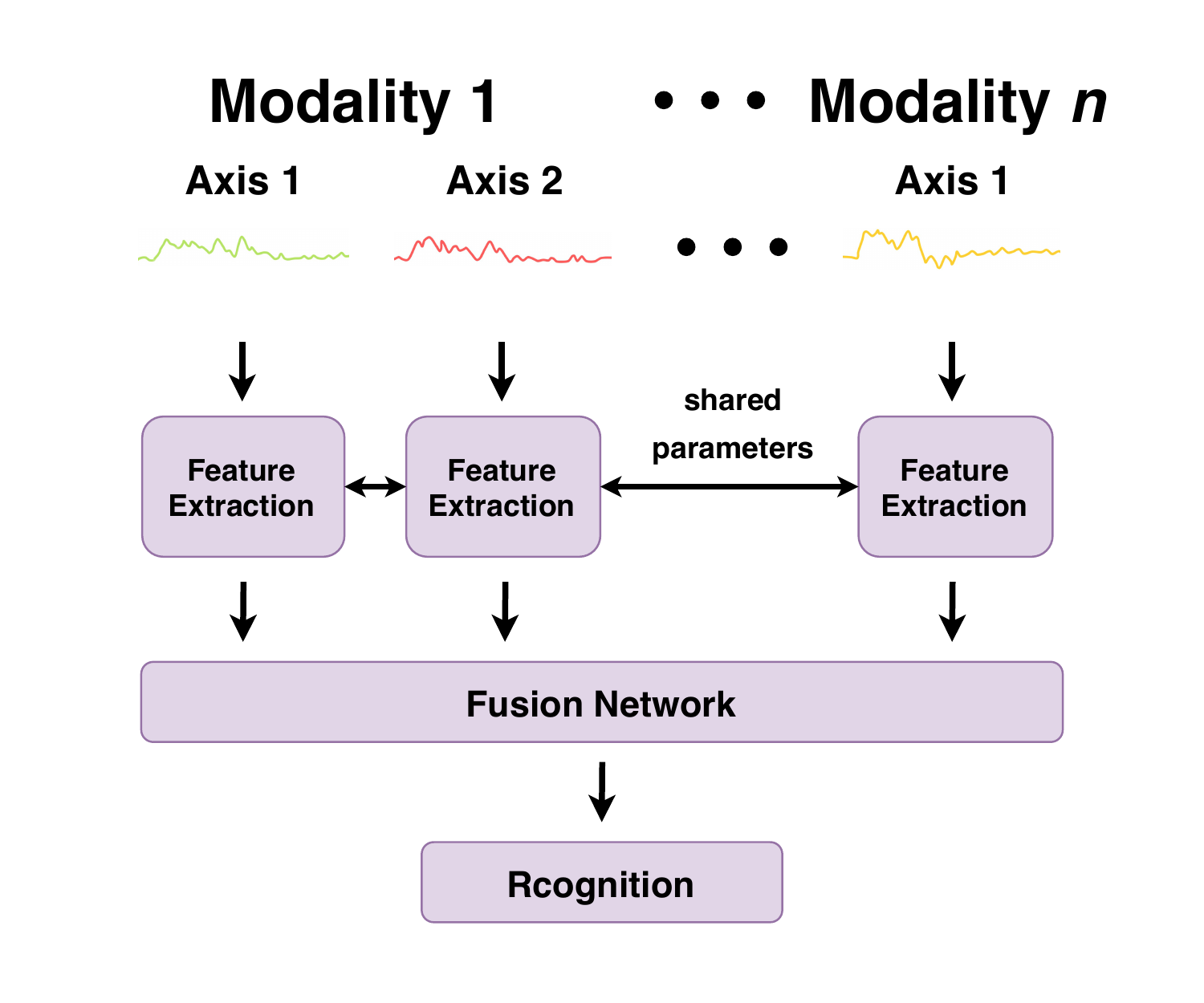}
    \label{fusion_share}}
\caption{Various strategies for feature fusion }
\label{fig:fusion}
\end{figure}

\paragraph{\textit{\textbf{Classifier Ensemble.}}} In addition to fusing features before interference, the integration of multiple modalities can be done by blending the recognition results from each modality as well. A range of ensemble approaches have been developed for fusing recognition results to yield an overall inference. For example, Guo et al. \cite{guo2016wearable046} proposed to use MLPs to create a base classifier for each sensing modality and incorporate all classifiers by assigning ensemble weights in the classifier level. When building the base classifiers, the authors not only considered the recognition accuracy but also emphasized the diversity of the base classifiers by inducing diversity measures. Thus, the diversity of different modalities is preserved, which is critical to overcoming the over-fit issues and to improving the overall generalization ability. Besides the conventional classifier ensemble, Khan et al. \cite{khan2017detecting} targeted the fall detection problem and introduced an ensemble of the reconstruction error from the autoencoder of each sensing modality.

The most attractive benefit of the classifier ensemble method is the scalability of additional sensors. A well-developed model of a specific sensing modality can be easily merged into an existing system by configuring the ensemble part only. Reversely, when a sensor is removed from a system, the recognition model can be freely adapted to this hardware change. Nevertheless, an intrinsic shortcoming of the ensemble fusion is that the inter-modality correlations may be underestimated due to the late fusion stage.

\subsubsection{Statistical Feature Extraction.} Different from deep learning-based feature extraction, feature engineering-based methods are able to extract meaningful features, such as statistical information. However, domain knowledge is usually required for manually designing such kind of features. In \cite{AAAI1816305}, a kernel embedding based solution is proposed to extract all statistical information of the activity data. However, spatial and temporal information is not considered in their model. Recently, Qian et al. \cite{qian2019novel128} managed to develop a Distribution-Embedded Deep Neural Network (DDNN) to integrate the statistical features with spatial and temporal information in an end-to-end deep learning framework for activity recognition. It encodes the idea of kernel embedding of distributions into a deep architecture, such that all orders of statistical moments could be extracted as features to represent each segment of sensor readings, and further combined with conventional spatial and temporal deep features for activity classification in an end-to-end training manner. 
The authors utilized an autoencoder to guarantee the injectivity of the feature mapping. They also introduced an extra loss function based on MMD distance to force the autoencoder to learn good feature representations of inputs. Extensive experiments on four datasets demonstrated the effectiveness of the statistical feature extraction methods. Although extracting statistical features has been explored in a deep-learning-based way, more reasonable and meaningful explanations on the extracted features are still undeveloped.

The technologies for feature extraction have their strengths and weaknesses. A summary of the advantages and limitations of different technologies is presented in Table \ref{tab:fe}.

\begin{table}[ht!]
\scriptsize
\centering
\caption{Advantages and Limitations of Different Works for Feature Extraction Approaches}
\scalebox{0.96}{
\begin{tabular}{|l|l|l|l|l|}
\hline
\textbf{Feature extraction} & \textbf{Approach} & \textbf{References}& \textbf{Advantages}                                                                                             & \textbf{Limitations}                                                   \\ \hline
\multirow{7}{*}{Temporal feature}    & mean/variance & \cite{vepakomma2015wristocracy} & -simple & \begin{tabular}[c]{@{}l@{}} -coarse\\ -unsatisfactory performance\end{tabular} \\ \cline{2-5} 
                                                & time-frequency &\cite{fan2018tagfree054}\cite{jiang2015human121}\cite{laput2019sensing018}& -capture frequency features & \begin{tabular}[c]{@{}l@{}}-experience dependent \end{tabular} \\ \cline{2-5} 
                                                & temporal CNN & \begin{tabular}[c]{@{}l@{}} \cite{duffner20143d119}\cite{grzeszick2017deep122}\cite{ha2016convolutional058}\cite{lee2017human115}\cite{ronao2015deep057}\\\cite{ronao2016human073}\cite{shen2018sam093}\cite{xi2018deep061}\cite{yang2015deep017}\cite{bai2020prototype} \end{tabular}& \begin{tabular}[l]{@{}l@{}}-capture local temporal \\\hspace{.5mm}features\end{tabular} & \begin{tabular}[l]{@{}l@{}}-limited in extracting global \\\hspace{.5mm}temporal features \end{tabular} \\ \cline{2-5} 
                                                & RNN &\begin{tabular}[c]{@{}l@{}}\cite{chen2016lstm}\cite{greff2016lstm}\cite{guan2017ensembles051}\cite{ishimaru2017towards026}\cite{yao2017deepsense047}\end{tabular} &\begin{tabular}[l]{@{}l@{}}-capture global temporal \\\hspace{.5mm}features\end{tabular}& -pre-slicing required \\ \cline{2-5} 
                                                & deep hybrid & \begin{tabular}[l]{@{}l@{}}\cite{gumaei2019hybrid}\cite{ordonez2016deep111}\cite{xu2019innohar071} \cite{yuki2018activity038}\\\cite{zou2018deepsense} \cite{wang2020push}\end{tabular}& \begin{tabular}[c]{@{}l@{}}-capture local and global \\ \hspace{.8mm}temporal features\end{tabular} & \begin{tabular}[c]{@{}l@{}}-complex structure\\ -high computation cost\end{tabular}   \\ \hline
\multirow{9}{*}{Multimodal feature}  & early fusion      & \begin{tabular}[c]{@{}l@{}}\cite{gu2018locomotion075}\cite{ha2017activity102}\cite{ha2015multi110}\cite{jiang2015human121}\cite{lee2017human115}\\\cite{li2016deep037}\cite{xi2018deep069}\end{tabular}  & -simple                                                                                                                      & \begin{tabular}[c]{@{}l@{}}-coarse\\ -unstable performance\end{tabular}  \\ \cline{2-5} 
                                                & sensor-based fusion     &\begin{tabular}[c]{@{}l@{}}\cite{choi2018confidence041}\cite{ha2016convolutional058}\cite{jiang2015human121}\cite{radu2018multimodal049}\cite{yao2017deepsense047}\end{tabular}&\begin{tabular}[c]{@{}l@{}} -capture sensor variance \\ -hierarchical features \end{tabular} & \begin{tabular}[c]{@{}l@{}}-limited in capturing \\\hspace{.5mm}intra-sensor variance \end{tabular}\\ \cline{2-5} 
                                                & axis-based fusion     &    \begin{tabular}[c]{@{}l@{}}   \cite{choi2018confidence041}\cite{ha2016convolutional058}\cite{zeng2014convolutional}\cite{zheng2014time117}\end{tabular}& \begin{tabular}[c]{@{}l@{}}-capture axis variance \\ -hierarchical features\end{tabular}  & \begin{tabular}[c]{@{}l@{}}-complex structure\\ -high computation cost\end{tabular}   \\ \cline{2-5} 
                                                & shared-filter fusion   & \cite{duffner20143d119}\cite{yang2015deep017}\cite{zebin2016human106}& \begin{tabular}[c]{@{}l@{}}-relative simple\\ -hierarchical features\end{tabular}                                                 & \begin{tabular}[c]{@{}l@{}}-limited in handling complex\\\hspace{.5mm} axis diversity \end{tabular}                                     \\ \cline{2-5} 
                                                & classifier ensemble & \cite{guo2016wearable046}\cite{khan2017detecting}& -high scalability  & \begin{tabular}[c]{@{}l@{}}-non end-to-end manner\\ -complex structure and training\end{tabular}   \\ \hline
Statistical feature      &  \multicolumn{1}{c|}{-}  & \cite{qian2019novel128} & -good interpretability   & \begin{tabular}[c]{@{}l@{}}-domain knowledge required\end{tabular}  \\ \hline
\end{tabular}
}
\label{tab:fe}
\end{table}

\subsection{Annotation Scarcity}
Section ~\ref{feature extraction} surveys the recent supervised deep learning methods for extracting distinguishable features from sensory data. One main characteristic of supervised learning methods is the necessity of a mass of labeled data to train the discriminative models. However, there may be some missing readings due to hardware issues making the sensor data temporally sparse that requires a specific structure of neural network to resolve \cite{AbedinRSR19}. Furthermore, it is more challenging to assign labels to a large amount of data. Firstly, the annotation process is expensive, time-consuming, and very tedious. Secondly, labels are subject to various sources of noise, such as sensor noise, segmentation issues, and the variation of activities across different people, which makes the annotation process error-prone. Therefore, researchers have begun to investigate unsupervised learning and semi-supervised learning approaches to reduce the dependence on massive annotated data.

\subsubsection{Unsupervised Learning}
Unsupervised learning is mainly used for exploratory data analysis to discover patterns among data. In \cite{li2011incorporating}, the authors examined the feasibility of incorporating unsupervised learning methods in activity recognition, 
but the community of activity recognition still needs more effective methods to deal with the high-dimensional and heterogeneous sensory data for activity recognition. 

Recently, deep generative models including Deep Belief Networks (DBNs) and autoencoders have become dominant for unsupervised learning. DBNs and autoencoders are composed of multiple layers of hidden units. They are useful in extracting features and finding patterns in massive data. Also, deep generative models are more robust against overfitting problems as compared to discriminative models \cite{mohamed2011acoustic}. So, researchers tend to use them for feature extraction to exploit unlabeled data as it is easy and cheap to collect unlabeled activity datasets.
According to Erhan et al. in \cite{erhan2010does}, a generative pretraining of a deep model guides the discriminative training to better generalization solutions. Pretraining a deep network on large-scale unlabeled datasets in an unsupervised fashion thus became very common.
The whole process for recognition can be divided into two parts. Firstly, the input data are fed to feature extractors, which are usually deep generative models, for pretraining, in order to extract features. Secondly, a top-layer or other classifier is added and then trained with labeled data in a supervised fashion for classification. During the supervised training, weights in the feature extractor may be fine-tuned.
For example, DBN-based activity recognition models are implemented in \cite{alsheikh2016deep007}. The unsupervised pretraining is followed by fine-tuning the learned weights in an up-down manner with available labeled samples.
In \cite{hammerla2015pd009}, the same pretraining process is conducted, but Restricted Boltzmann Machines (RBMs) are applied to learn a generative model of the input features.
In another work \cite{plotz2011feature010}, Pl{\"o}tz et al. proposed to use autoencoders for unsupervised feature learning as an alternative to Principal Component Analysis (PCA) for activity recognition in ubiquitous computing. And the authors in \cite{chikhaoui2017towards020,gu2018locomotion075,zeng2017semi104} employed the variants of autoencoders such as stacked autoencoders \cite{chikhaoui2017towards020}, stacked denoising autoencoders \cite{gu2018locomotion075}, and CNN autoencoders \cite{zeng2017semi104} to combine automatic feature learning and dimensionality reduction in one integrated neural network for activity recognition. 
In a recent work \cite{bai2019motion2vector130}, Bai et al. proposed a method called Motion2Vector to convert a time period of activity data into a movement vector embedding within a multidimensional space. To fit with the context of activity recognition, they use a bidirectional LSTM to encode the input blocks of the temporal wrist-sensing data. 

Despite the success of deep generative models in unsupervised learning for human activity recognition, unsupervised learning still cannot undertake the activity recognition tasks independently since unsupervised learning is not capable of identifying the true labels of activities without any labeled samples presenting the ground truth. 
Therefore, the aforementioned methods can be considered as semi-supervised learning, in which both labeled data and unlabeled data are leveraged for training the neural networks. 

\subsubsection{Semi-supervised Learning}

\begin{figure}[ht!]
\setlength {\belowcaptionskip} {-0.4cm}
\centering
\subfloat[Co-training]{
    \includegraphics[width=.4\textwidth]{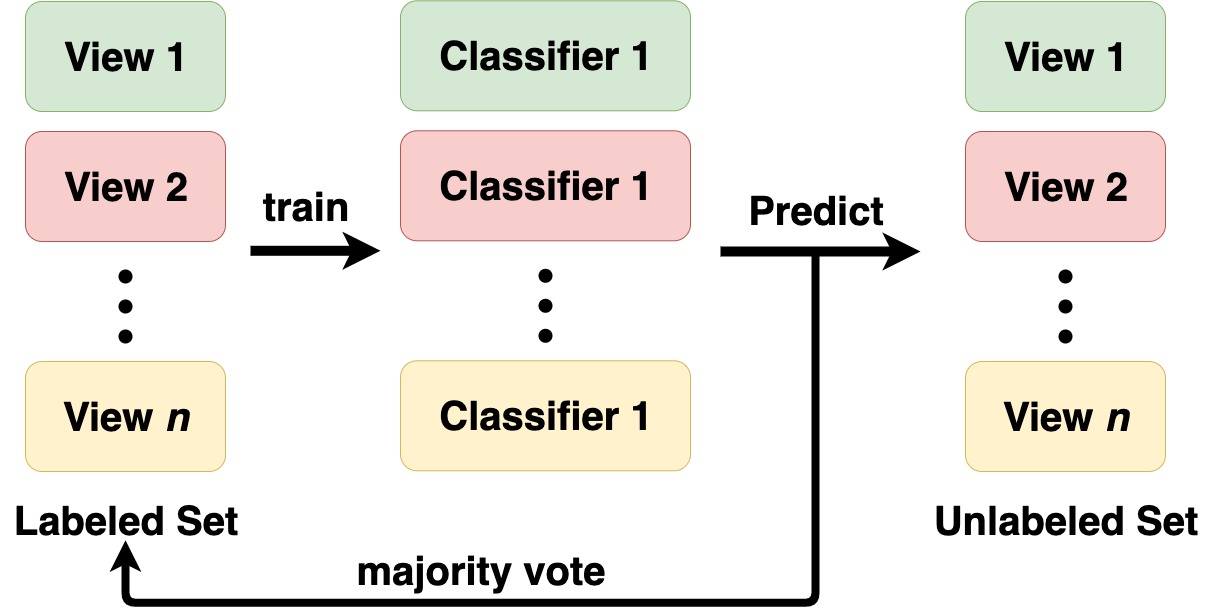}
    \label{scarcity-cotraining}}
    \hspace{1cm}
\subfloat[Active Learning]{
    \includegraphics[width=.4\textwidth]{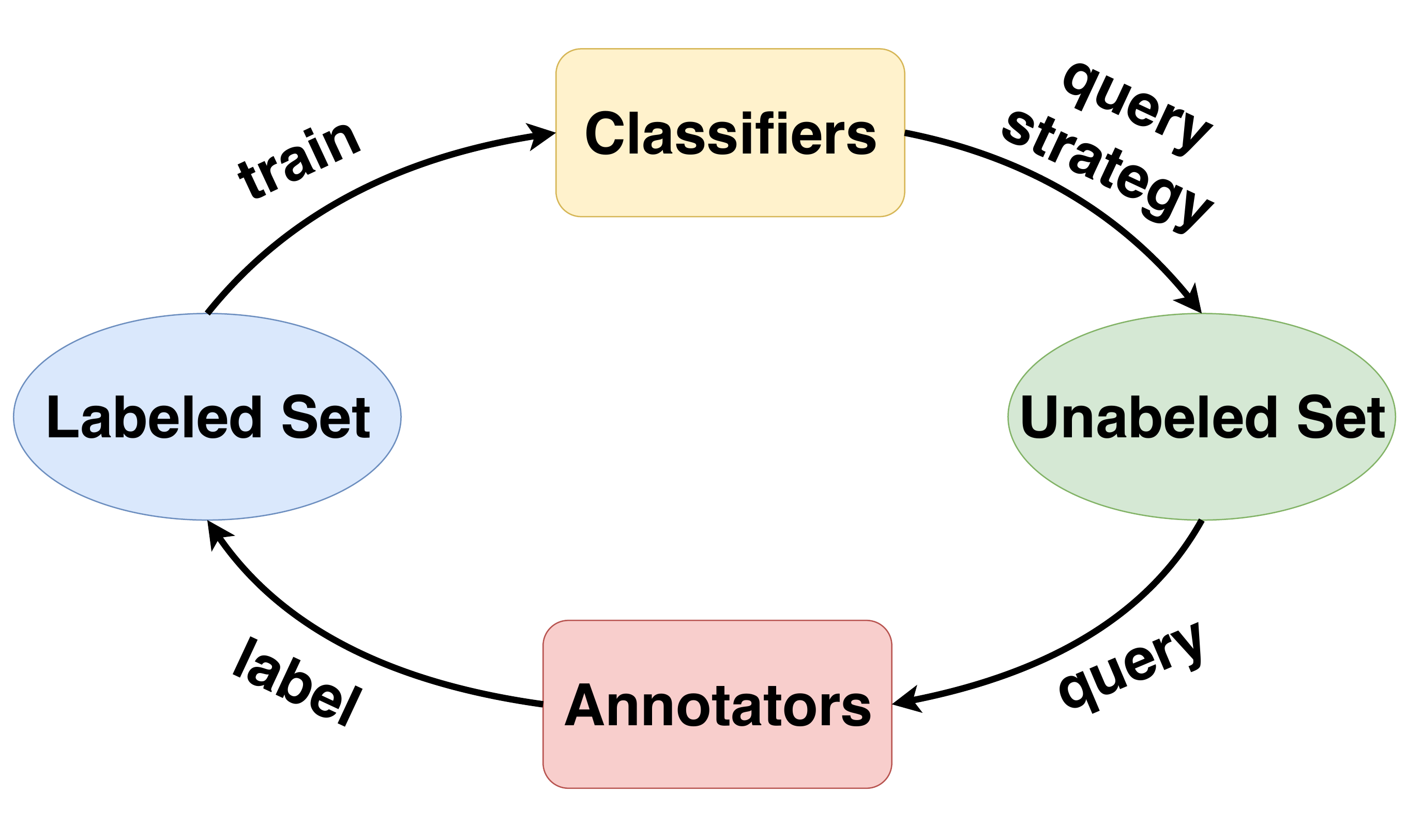}
    \label{scarcity-activelearning}}
\caption{Co-training and active learning for Annotation Scarcity}
\label{fig:scarcity}
\end{figure}
Semi-supervised learning has shown a growing trend in activity recognition because of the difficulty in obtaining labeled data \cite{yao2016learning}. A semi-supervised method requires less labeled data and massive unlabeled data for training. How to utilize unlabeled data for reinforcing the recognition system has become a point of interest. Some works have explored to promote classic semi-supervised learning methods on activity recognition, such as manifold learning \cite{qian2019distribution,ma2019labelforest}. Recently, as deep learning is powerful in capturing patterns from data, various semi-supervised learning has been incorporated for activity recognition such as co-training, active learning, and data augmentation.

\textbf{\textit{Co-training}} was proposed by Blum and Mitchell in 1998 \cite{blum1977combining}. It was an extension of self-learning. In self-learning approaches, a weak classifier is first trained with a small amount of labeled data. This classifier is used for classifying the unlabeled samples. The samples with high confidence can be labeled and added to the labeled set for re-training the classifier.
In co-training, multiple classifiers are employed, each of which is trained with one individual view of training data. Likewise, the classifiers select unlabeled samples to add to the labeled set by confidence score or majority voting. The whole process of co-training can be seen in Figure~\ref{fig:scarcity} (a). With the training set augmented, the classifiers are enhanced.
Blum and Mitchell \cite{blum1977combining} suggested that co-training is fully effective under three conditions: (a) multiple views of training data are not strongly correlated, (b) each view contains sufficient information for learning a weak classifier, (c) the views are mutually redundant. In respect of sensor-based human activity recognition, co-training is compatible because multiple modalities can be regarded as multiple views. Chen et al. \cite{chen2019semisupervised125} applied co-training with multiple classifiers on different modalities of the data. Three classifiers are trained on acceleration, angular velocity, and magnetism, respectively. The learned classifiers are used for predicting the unlabeled data after each training round. If most of the classifiers reach an agreement on predicting an unlabeled sample, this sample is labeled and moved to the labeled set for the next training round. The training flow is repeated until no confident samples can be labeled, or the unlabeled set is empty. Then a new classifier is trained on the final labeled set with all modalities.

Co-training is like human learning. People can learn new knowledge from existing experience, and new knowledge can be used to summarize and accumulate experience. Experience and knowledge constantly interact with each other. Similarly, co-training uses current models to select new samples that they can learn from, and the samples help to train the models for the next selection. However, automatic labeling may introduce errors. Acquiring correct labels can improve accuracy.

\textbf{\textit{Active learning}} is another category in semi-supervised learning. Different from self-learning and co-training which label the unlabeled samples automatically, active learning requires annotators who are usually experts or users to label the data manually. In order to lighten the burden of labeling, the goal of active learning is to select the most informative unlabeled instances for annotators to label and improve the classifiers with these data so that minimal human supervision is needed.
Here the most informative instances denote the instances that bring the most enormous impact on the model if their labels are available.
A general framework of active learning can be seen in Figure~\ref{fig:scarcity} (b). It includes a classifier, a query strategy, and an annotator. The classifier learns from a small amount of labeled data, selects one or a set of the most useful unlabeled samples via query strategy, ask the annotator for true labels, and utilize the new labels for further training and next query. The active learning process is also a loop. It stops when it meets the stop criteria. There are two common query strategies for selecting the most profitable samples which are uncertainty and diversity. Uncertainty can be measured by information entropy. Larger entropy means higher uncertainty and better informativeness. Diversity means that the queried samples should be comprehensive, and the information provided by them are non-repetitive and non-redundant. In \cite{stikic2008exploring}, the authors applied two query strategies. One of them is to select samples with lowest prediction confidence, and the other one resort to the idea of co-training, but it oppositely selects samples with high disagreement among classifiers.

Deep active learning approaches are deployed in activity recognition \cite{hossain2018deactive053,hossain2019active131}. Hossain et al. \cite{hossain2018deactive053} considered that traditional active learning methods merely choose the most informative samples which only occupy a small fraction of the available data. In this way, a large number of samples are discarded. Although the selected samples are vital for training, the discarded samples are also of value on account of the substantial amount. Therefore, they proposed a new method to combine active learning and deep learning in which not only the most informative unlabeled samples are queried but the less necessary samples are also leveraged. The data is first clustered with K-means clustering. While the intuitive idea is to query the optimal samples such as the centroids of the clusters, in this work, the neighboring samples are also queried. The experiments show that the proposed method can achieve the optimal results by labeling 10\% of the data.

Hossain and Roy \cite{hossain2019active131} further investigated two problems of deep active learning and human activity recognition. The first problem is that outliers can be easily mistaken for important samples. When entropy is calculated for selection, apart from informativeness, larger entropy may also mean outliers because outliers belong to none of the classes. Therefore, a joint loss function was proposed in \cite{hossain2019active131} to address this problem. Cross-entropy loss and information loss are jointly minimized to reduce the entropy of outliers. The second problem considered in this work is how to reduce the workload of annotators as annotators are required to master domain knowledge for accurate labels. Multiple annotators are employed in this work. They are selected from the intimate people of users. The annotator selection is made by the reinforcement learning algorithm according to the discrepancy and the relations of users. The contextual similarity is used to measure the relations among users and annotators. The experimental results show that this work has an 8\% improvement in accuracy and has a higher convergence rate.

Co-training and active learning are based on the same idea of rebuilding the model upon labels of unlabeled data. Data augmentation with synthesizing new activity data is another way when data collection is challenging in specific scenarios such as resource-limited or high-risk scenarios.

\textbf{\textit{Data augmentation}} with synthesizing data indicates generating massive fake data from a small amount of real data so the fake data can facilitate to train the models. One popular tool is Generative Adversarial Network (GAN).
GAN was firstly introduced in \cite{goodfellow2014generative}. GAN is powerful in synthesizing data that follow the distribution of training data. A GAN is composed of two parts, a generator and a discriminator. The generator creates synthetic data and the discriminator evaluates them for authenticity. The goal of the generator is to generate data that are genuine enough to cheat the discriminator while the goal of the discriminator is to identify images generated by the generator as fake. The training is in an adversarial way, which is based on a min-max theory. During training, the generator and the discriminator mutually improve their performance in generation and discrimination. Variants of GANs has been applied to different fields such as language generation \cite{press2017language} and image generation \cite{zhu2019learning}. 

The first work about data augmentation with synthesizing sensory data for activity recognition is called SensoryGANs \cite{wang2018sensorygans064}. As sensory data is heterogeneous, a unified GAN may not be enough to depict the complex distribution of  different activities. Wang et al. employed three activity-specific GANs for three activities. After generation, the synthetic data are fed into classifiers for prediction with original data. We should note that although this work uses deep generative networks, the generation process depends on labels so the process is not unsupervised. 
Zhang et al. \cite{zhang2019adversarial124} proposed to use semi-supervised GAN for activity recognition. Different from regular GAN, the discriminator in semi-supervised GAN makes a $K + 1$ class classification that includes activity classification and fake data identification. To ensure the distribution of the generated data to trend to the authentic distribution, a prearranged distribution is provided as inputs by Variational AutoEncoders (VAEs) instead of Gaussian noises. The aim of VAEs is to provide distributions that represent the distributions of input data. Moreover, VAE++ was proposed to guarantee that the inputs are exclusive for each training sample. Overall, the unified framework combining VAE++ and semi-supervised GAN proves to be effective in activity recognition.

Table ~\ref{tab: annotation scarcity} summarizes recent deep learning works for annotation scarcity in activity recognition and their advantages and disadvantages.

\begin{table}[ht!]
\scriptsize
\centering
\caption{Advantages and Limitations of Different Works for Annotation Scarcity}
\begin{tabular}{|l|l|l|l|l|}
\cline{1-5}
\begin{tabular}[c]{@{}c@{}}\textbf{Training scheme}\end{tabular} & \textbf{Approach} & \textbf{References} & \textbf{Advantages} & \textbf{Limitations} \\ \cline{1-5}
Unsupervised & pretraining & \begin{tabular}[l]{@{}l@{}}\cite{alsheikh2016deep007}\cite{bai2019motion2vector130}\cite{chikhaoui2017towards020}\cite{gu2018locomotion075}\\\cite{hammerla2015pd009}\cite{plotz2011feature010}\cite{zeng2017semi104}\end{tabular} & \begin{tabular}[l]{@{}l@{}} -feature learning without \\ \hspace{.5mm} labels \end{tabular} & \begin{tabular}[l]{@{}l@{}} -rely on ground truth for training  \\ \hspace{.5mm} activity classifiers \end{tabular}\\ \cline{1-5}
\multirow{4}{*}{Semi-supervised} & co-training & \cite{chen2019semisupervised125} & \begin{tabular}[l]{@{}l@{}} -use both labeled and \\ \hspace{.5mm} unlabeled data \\ -assign labels to unlabeled \\ \hspace{.5mm} data automatically\end{tabular} & \begin{tabular}[l]{@{}l@{}} -at least two data modalities \\ \hspace{.5mm} required \\ -need training multiple classifiers \\\hspace{.5mm} each iteration\end{tabular}\\ \cline{2-5}

& active learning & \begin{tabular}[l]{@{}l@{}} \cite{hossain2018deactive053}\cite{hossain2019active131}\end{tabular} & \begin{tabular}[l]{@{}l} -high labeling efficiency and \\ \hspace{.5mm}accuracy \end{tabular} & \begin{tabular}[l]{@{}l@{}} -human labeling required \end{tabular} \\ \cline{2-5}

& \begin{tabular}[l]{@{}l@{}}data augmentation\end{tabular} & \begin{tabular}[l]{@{}l@{}} \cite{wang2018sensorygans064}\cite{zhang2019adversarial124}\end{tabular} &\begin{tabular}[l]{@{}l@{}} -enhance model generalization \end{tabular} & \begin{tabular}[l]{@{}l@{}} -make less use of unlabeled data\end{tabular}\\ \cline{1-5}
\end{tabular}
\label{tab: annotation scarcity}
\end{table}

\subsection{Class Imbalance}
The primary contributor to the success of deep learning technique is the availability of a large volume of training data due to modern information technology. Most existing research on human activity recognition follows a supervised learning manner, which requires a significant amount of labeled data to train a deep model. However, some sensor data of specific activities are challenging to obtain, such as those related to falls of elderly people. In addition, raw data recorded from unconstrained conditions is naturally class-imbalanced. When using an imbalanced dataset, conventional models tend to predict the class with the majority number of training samples while ignoring the class with few available training samples. Therefore, it is urgent to determine the class imbalance issue for developing an effective activity recognition model. Methods of dealing with class imbalance can be divided into two groups.

\subsubsection{Data Level}
The most intuitive path to tackling the imbalance problem is to re-sample the class with the largest number of samples \cite{alani2020classifying}. However, such a method is at the risk of reducing the total amount of training samples and omitting some critical samples with featured characteristics. 
In contrast, augmenting new samples to the class with a minority number of samples could not only keep all original samples but also enhance models' robustness. Grzeszick et al. \cite{grzeszick2017deep122} utilized two augmentation methods, Gaussian noises perturbation and interpolation, to tackle the problem of class imbalance. The augmentation approaches could preserve the coarse structure of the data, but a random time jitter in the sensor’s sampling process is simulated. They created a larger number of samples for the under-represented classes and ensure that each class has at least a certain percentage of data in the training set. 

\subsubsection{Algorithmic Level}
Another direction of solving the imbalance concern is to modify the model-building strategy instead of directly balancing the training dataset. 
In \cite{guan2017ensembles051}, Guan and Pl{\"o}tz utilized the $F$1-score rather than the conventional cross-entropy as the loss function to address the imbalance problem. Because the $F$1-score considers both the recall and precision aspects, classes with different numbers of training samples are equally taken into account. Besides the class imbalance of original datasets, it is also a non-negligible problem for a semi-supervised framework as the process of gradually labeling unlabeled samples may create uneven new numbers of labels across different classes. Chen et al. \cite{chen2019semisupervised125} concerned class imbalance in small labeled datasets. They leveraged a semi-supervised framework, co-training, to enrich the labeled set in cyclic training rounds. To balance the training samples across classes while simultaneously maintain the distributions of the samples, a pattern-preserving strategy was proposed before the training phase of the co-training framework. K-means clustering was first adopted to mine latent activity patterns of each activity. Then, sampling is applied to each pattern. The main goal is to guarantee that the numbers of all the patterns of all activities are even. A summary of the advantages and limitations of different works for resolving class imbalance is presented in Table \ref{tab: ci}.

\begin{table}[ht!]
\scriptsize
\centering
\caption{Advantages and Limitations of Different Works for Class Imbalance}
\begin{tabular}{|l|l|l|l|l|}
\cline{1-5}
\textbf{Balancing scheme} & \textbf{Approach} & \textbf{References} & \textbf{Advantages} & \textbf{Limitations} \\ \cline{1-5}
\multirow{3}{*}{Data level} & re-sampling & \cite{alani2020classifying} & \begin{tabular}[l]{@{}l@{}} -simple balancing process \\ -free of noises \end{tabular} & \begin{tabular}[l]{@{}l@{}} -decrease the amount of sample \\ -may miss featured samples \end{tabular} \\ \cline{2-5} 
& augmentation & \cite{grzeszick2017deep122} & \begin{tabular}[l]{@{}l@{}} -enhance model robustness \\ -keep all recording samples\end{tabular} & -may induce unexpected noises\\ \cline{1-5}
Algorithmic level & \multicolumn{1}{c|}{-} & \cite{chen2019semisupervised125}\cite{guan2017ensembles051} & \begin{tabular}[l]{@{}l@{}} -free of data preprocess \\ -keep all recording samples\end{tabular} & \begin{tabular}[l]{@{}l@{}} -not generic \\ -careful parameter tuning required \end{tabular}\\ \cline{1-5}
\end{tabular}
\label{tab: ci}
\end{table}

\subsection{Distribution Discrepancy}
Many state-of-the-art approaches for human activity recognition assume that the training data and the test data are independent and identically distributed (i.i.d.). However, this is impractical since there is distribution discrepancy between training data and test data in activity recognition. The distribution discrepancy in sensory data can be divided into three categories by reason. The first one is the discrepancy between users which stems from different motion patterns when activities are performed by different people. 
The second discrepancy is with time. In a dynamic streaming environment, data distributions of activities are changing over time, and new activities may also emerge.
The third category is the discrepancy in sensors.
Sensors used for human activity recognition are usually sensitive. A small variation in sensors can cause a significant disturbance in the sensory data.
The factors that may potentially bring about discrepancy with sensors include sensor instances, types, positions, and layouts in the environment.
We can also categorize the discrepancy into homogeneous discrepancy and heterogeneous discrepancy by character \cite{day2017survey}. In homogeneous discrepancy, training data and test data have the same attributes and the same feature spaces. In heterogeneous discrepancy, the feature space of training data and test data may differ in dimensions or attributes.
Typically, the discrepancy among users and time belongs to homogeneous discrepancy while the discrepancy with the number of sensor instances, sensor types, and sensor layouts is heterogeneous as these factors may cause change in attributes and dimensions.
The following section summarizes the literature by reason (i.e., users, time, and sensors), but the perspective of homogeneous and heterogeneous discrepancy is also inspiring.


Before taking a closer look at the factors that cause distribution discrepancy in sensory data, we briefly introduce \textit{transfer learning} \cite{pan2009survey}. Transfer learning is a common machine learning technique that transfers the classification ability of the learning model from one predefined setting to a dynamic setting. Transfer learning is particularly effective in solving distribution discrepancy problems. It avoids the decline in the performance of learning models when the training data and the test data follow different distributions. In the activity recognition context, this problem appears when activity recognition models are deployed for application in a different configuration with where they are trained. In transfer learning, \textit{source domain} refers to domains that contain massive annotated data and knowledge, and the goal is to leverage the information from the source domain to annotate the samples in the \textit{target domain}. Regarding activity recognition, the source domain corresponds to the original configuration, and the target domain denotes the new deployment that the system has never encountered (e.g., new activities, new users, new sensors). In the following sections, we detailedly introduce three categorizes of discrepancy and how the state-of-the-art approaches manage to mitigate the discrepancy. Most of them are based on transfer learning.

\subsubsection{Distribution Discrepancy with Users}

Owing to biological and environmental factors, the same activity can be performed differently by different individuals. For example, some people walk slowly and some prefer to walk faster and more dynamically. Since people have diverse behavior patterns, data from different users are distributed variously.
Usually, if the models are trained and tested with data that are collected from a specific user, the accuracy can be rather high. However, this setting is impractical.
In practical human activity recognition scenarios, while a certain number of participants' data can be collected and annotated for training, the target users are usually unseen by the systems. So the distribution divergence between the training data and the test data appears as a challenge in human activity recognition, and the performance of the models falls dramatically across users. The research on personalized models for a specific user is significant.
Recently, personalized deep learning models for distribution discrepancy among users in activity recognition have been explored. Woo et al. \cite{woo2016rnn100} proposed an approach to build an RNN model for each individual. Learning Hidden Unit Contributions (LHUC) were applied in \cite{matsui2017user103} where a particular layer with few parameters is inserted between every two hidden layers of CNN, and the parameters are trained using a small amount of data. Rokni et al. \cite{rokni2018personalized005} proposed to personalize their models with transfer learning. In the training phase, CNN is firstly trained with data collected from a few participants (source domain). In the test phase, only the top layers of the CNN are fine-tuned with a small amount of data for the target users (target domain). Annotation for target users is required.
GAN is also serviceable for addressing distribution discrepancy among users. In \cite{soleimani2019cross088}, the authors generated data of the target domain directly from the source domain with GANs to enhance the training of the classifier. Chen et al. \cite{chen2019distributionally077} further defined person-specific discrepancy and task-specific consistency for people-centric sensing applications. 
Person-specific discrepancy means the distribution divergence of data collected from different people, and task-specific consistency denotes the inherent similarity of the same activity.
They proved that reducing person-specific discrepancy and preserving task-specific consistency guarantee the recognition accuracy after transferring. 
\cite{chen2020metier} combines activity recognition and user recognition with a multi-task model. The proposed method shares parameters between the activity module and the user module so the activity recognition performance can be boosted by features learned from the user recognition module. To transfer important knowledge between the two modules, a mutual attention mechanism is deployed.

\begin{figure}[ht!]
\setlength {\belowcaptionskip} {-0.5cm}
\centering
\subfloat[Concept Drift]{
    \includegraphics[width=.3\textwidth]{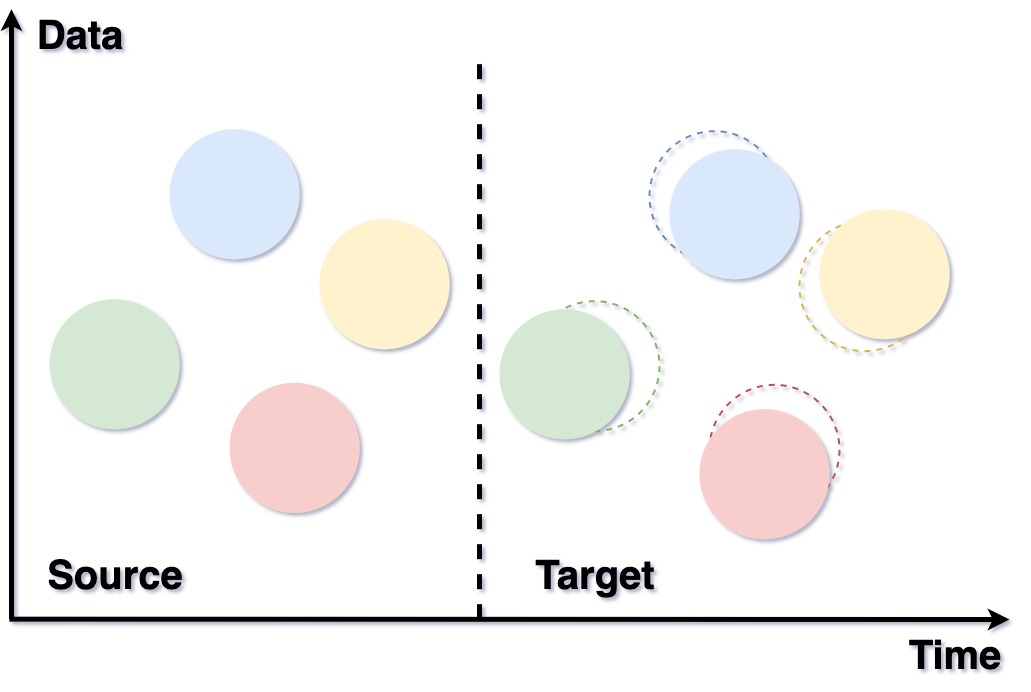}
    }
\subfloat[Concept Evolution]{
    \includegraphics[width=.3\textwidth]{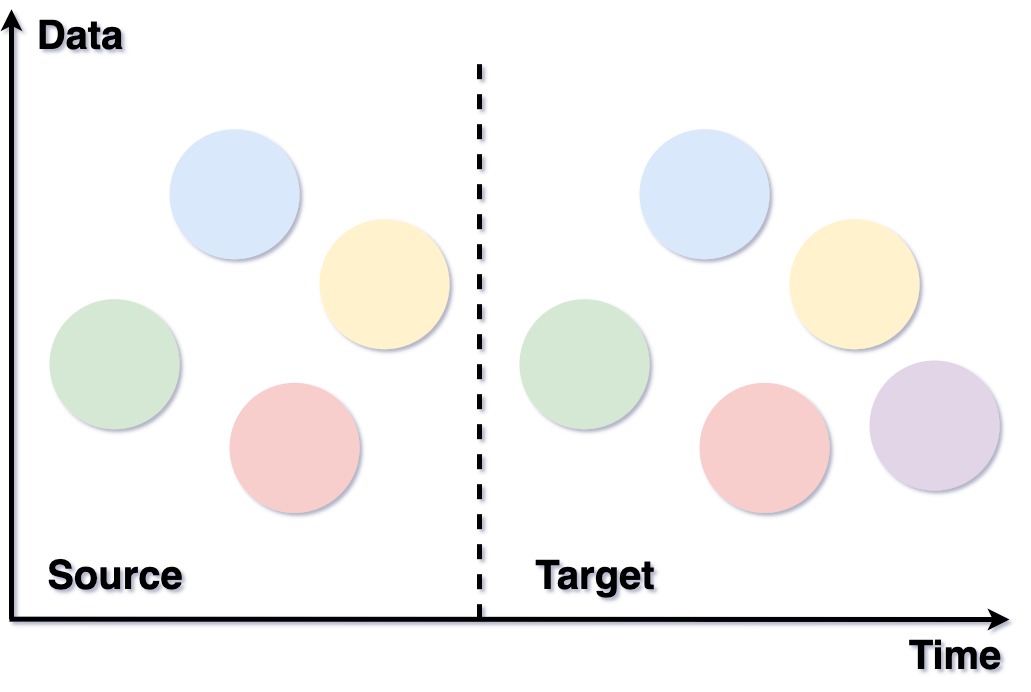}
    }
\subfloat[Open-Set]{
    \includegraphics[width=.3\textwidth]{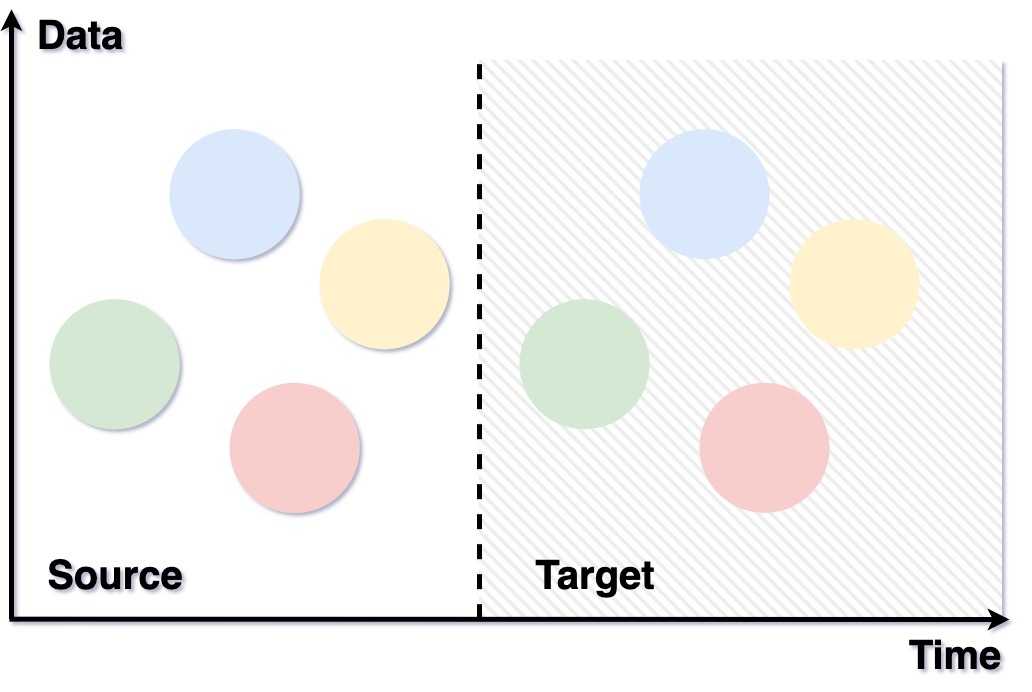}
    }   
\caption{Distribution Discrepancy with Time}
\label{fig:concept}
\end{figure}

\subsubsection{Distribution Discrepancy with Time}

Human activity recognition systems collect dynamic and streaming data that logs people's motions. In a real-world recognition system, the initial training data that portrays a set of activities is collected to train an original model, then the model is configured for future activity recognition.
In long-term systems which are longer than months or even years, a natural feature that we should concern is that the streaming sensory data changes over time. Three problems can be derived from the distribution discrepancy with time in line with the extent of change and the extent of the need in recognizing the new concepts of data. They are the concept drift problem, the concept evolution problem, and the open-set problem.

\textit{\textbf{Concept Drift.}} Figure~\ref{fig:concept}(a) shows the first problem of distribution discrepancy with time in activity recognition called concept drift \cite{schlimmer1986incremental}. It denotes the distribution shift between the source domain and the target domain. Concept drift can be abrupt or gradual \cite{abdallah2018activity001}. To accommodate the drift, deep learning models should incorporate \textit{incremental training} to continuously learn new concepts of human activities from newly coming data. For example, an ensemble
classifier termed multi-column bi-directional LSTM was proposed in \cite{tao2016multicolumn074}. The model leverages new training samples gradually via incremental learning. 
\textit{Active learning} is a special type of incremental learning. In streaming data systems, active learning  queries ground truth for samples when change is detected. It encourages to select the most efficient samples to update the models for the new concepts. That is why active learning can facilitate deep learning models to mitigate the discrepancy with time of the streaming sensory data \cite{gudur2019activeharnet081,saeedi2017closed105}. In this way, Gudur et al. \cite{gudur2019activeharnet081} proposed a deep Bayesian CNN with dropout to obtain the uncertainties of the model and select the most informative data points to be queried according to the uncertainty query strategy. Owing to the active learning, the model supports updating continuously and capturing the changes of data over time.

\textit{\textbf{Concept Evolution.}}
Figure~\ref{fig:concept}(b) represents the distribution of concept evolution. Concept evolution denotes the emergence of new activities in the streaming data. The appearance of concept evolution is because collecting labeled data for all kinds of activities in the initial learning phase is impractical.
Firstly, despite the effort, the initial training set in an activity recognition system is only able to contain a limited number of activities. 
Secondly, people can perform new activities that they never did before the initial training of the activity recognition system (e.g., learning to play guitar for the first time).
Thirdly, it is difficult to collect some certain activities such as people falling down. However, these activities still may appear in the test or the application phase. 
Thus, in the application phase, the concepts of the new activities still need to be learned. It is essential to study activity recognition systems which can recognize new activities in the streaming data settings. Nevertheless, this is difficult due to the restricted access to annotated data in the application phase. 
One approach is to decompose activities into \textit{mid-level features} such as arm up, arm down, leg up, and leg down. This method demands experts to define the mid-level attributes for further training, and the capability is limited when new activities composed of new attributes appear \cite{nair2019attrinet132}.
Other deep learning methods for activity concept evolution are still less explored, so some researchers take a step back and study the problem of open-set.

\textit{\textbf{Open-Set.}}
Open-set problem is currently a trending topic. Before that, most of the state-of-the-art works are for ``closed-set'' problems where the training set and the test set contain the same set of activities.
Open-set also originates from the fact that we can never collect sufficient kinds of activities in the initial training phase. But compared with concept evolution problems, the solutions to open-set problems only need to identify whether the test samples belong to the target activities, rather than exactly recognize the activities. Figure~\ref{fig:concept}(c) represents the distribution of open-set problems where the shadow means the space where new activities may emerge. An intuitive solution to open-set problems is to build a negative set so that they can be considered in a closed-set way. A deep model based on GAN is proposed in \cite{yang2019open067}. The authors generate fake samples with GAN to construct the negative set, and the discriminator of the GAN can be seamlessly used as the open-set classifier.

\subsubsection{Distribution Discrepancy with Sensors}
Due to the sensitivity of sensors, a tiny variation in the sensors may lead to substantial changes in the data collected or transmitted by the sensors. The influential factors of sensors include the instances, types, positions, and layouts in the environment. To illustrate, instances of sensors may have different parameters such as the sampling rate; different types of sensors collect totally different types of data with varying shapes, frequencies, and scales; wearable sensors attached to positions of human body only record motions in the corresponding body parts; environmental layouts of device-free sensors influence the propagation of signals. All of these factors may cause drops in the recognition accuracy when the classifiers are not trained for specific device deployments. Therefore, seamless deep learning models for activity recognition in the wild is necessary. \cite{morales2016deep025} proves that features learned by deep learning models are transferable across sensor types and sensor deployments for activity recognition. 

\textbf{\textit{Sensor Instances.}}
Even when data is collected in the same setting, and only the sensor instances are different, for example, a person replaces his smartphone with a new one, the recognition accuracy still declines soon. Both the hardware and the software are responsible. In fact, owing to the imperfections in the production process, sensor chips show variation in the same conditions \cite{dey2014accelprint}. Also, the performance of devices differs in different software platforms \cite{blunck2013heterogeneity}. For example, APIs, resolutions, and other factors are all influential to the performance of sensors.
There have been a few works developing deep learning models to address distribution discrepancy problems caused by different sensor instances. One notable work is data augmentation with GANs \cite{mathur2018using_023}. Data augmentation is a solution of enriching training sets so that both the size and the quality of training sets meet the requirement of training a powerful deep learning model. A discrepancy generator that synthesizes heterogeneous data from different sensor instances under various degrees of disturbance is developed in \cite{mathur2018using_023}. The aim is to replenish the training set with sufficient discrepancy. Moreover, the authors deploy a discrepancy pipeline with two parameters that control the discrepancy of the training set.

\textbf{\textit{Sensor Types and Positions.}}
In this section, we introduce the distribution discrepancy of sensory data caused by different sensor types and positions on human bodies because these two factors usually appear together. Thanks to the pervasiveness of wearables sensors and IoT equipment, people can wear more than one smart devices to assist their daily life. And it is also common that users replace their smart devices or buy new electronic products.
Since some devices are based on the same platforms (e.g., iPhone and Apple Watch), people prefer the activity recognition system to seamlessly recognize activities that are observed by the new device with models trained with the old devices. 
In terms of positions, the devices should be attached to different body positions according to the types. For example, a smartwatch should be attached to the user's wrist while a smartphone can be put in a pocket of a trouser or shirt. It is obvious that devices on different body positions will lead to tremendous changes in their collected signals because the signals are stimulated by the motions of corresponding body parts. Therefore, there are two issues raised by such changes that urgently need to be considered to address the distribution discrepancy with sensor types and positions. Firstly, massive data from the new sensors or new positions is required so that the new distribution can be estimated rather completely. Secondly, most of the existing works still mediocrely characterize the old data and the new data with the same features, which is impractical when sensor types and positions are not fixed.
For instance, KL divergence is minimized between the parameters of CNNs which are trained by the old data and the new data, respectively in \cite{khan2018scaling036}. In order to address the issue mentioned, Akbari and Jafari \cite{akbari2019transferring_024} designed stochastic features that are not only discriminative for classification but also able to reserve the inherent structures of the sensory data. The stochastic feature extraction model is based on a generative autoencoder.

Wang et al. \cite{wang2018deep090} further posed a question about how to select the best source positions for transfer when there are multiple source positions available. This question is pragmatic since the smart devices can be placed in diverse positions such as on wrist, in a pocket, or on nose (e.g., goggles), and inappropriate selection may lead to negative transfer. \cite{gjoreski2019cross133} proves that the similarity between domains in transfer learning is determinative. \cite{wang2018deep090} suggests that higher similarity indicates better transfer performance between two domains. 
Therefore, Chen et al. \cite{chen2019cross089} assumed that data samples of the same activities are aggregated in the distribution space even when they are from different sensors. They propose a stratified distance which is class-wise to measure the distances between domains. Wang et al. \cite{wang2018deep090} proposed a semantic distance and a kinetic distance to measure domain distances, where the semantic distance involves spatial relationships between data collected from two positions and the kinetic information concerns the relationships of motion kinetic energy between two domains.

\textbf{\textit{Sensor Layouts and Environments.}}
Sensor layouts are in regard to device-free sensors such as WiFi and RFID. The signals collected by the receivers are usually considerably influenced by the layouts and the environments. The reason is that during the signals are transmitted, the signals are inevitably reflected, refracted, and diffracted by media and barriers such as air, glass, and walls. And the spatial positions of the receivers also play a role. Despite the maturity in building classification models for device-free activity recognition, very few works focus on how to get equally accurate recognition performance when sensors are configured in the wild. One example is \cite{jiang2018towards032}, where an adversarial network is incorporated with deep feature extraction models to remove the environment-specific information and extract the environment-independent features.

It should be noted that all the aforementioned methods need either labeled or unlabeled data from the target domain to update their models. In real world, a one-fits-all model that only requires one-time training and is general enough to fit all scenarios is indispensable. Zheng et al. \cite{zheng2019zero098} defined Body-coordinate Velocity Profile (BVP) to capture domain-independent features. The features represent power distributions over different velocities of body parts and are unique to individual activities. The experimental results show that BVP is advantageous in cross-domain learning, and it fits all kinds of domain factors including users, sensor types, and sensor layouts. One-fits-all is a new direction for researchers to mitigate the distribution discrepancy problem in activity recognition.

In conclusion, we review three categories of distribution discrepancy in activity recognition. They are caused by different users, time streaming, and sensor deployments. They are further categorized according to the extent of change or the main reason for changes.
Table \ref{tab: discrepancy} summarizes the advantages and limitations of different works for resolving distribution discrepancy in activity recognition.

\begin{table}[ht]
\scriptsize
\centering
\caption{Advantages and Limitations of Different Works for Distribution Discrepancy}
\label{tab: discrepancy}
\scalebox{0.96}{
\begin{tabular}{|l|l|l|l|l|}
\hline
\textbf{\begin{tabular}[l]{@{}l@{}}Discrepancy\\Type\end{tabular}} & \textbf{Approach}& \textbf{References}& \textbf{Advantages}& \textbf{Limitations}\\ \hline
\multirow{6}{*}{User} 
& user-specific models &\cite{woo2016rnn100}& \begin{tabular}[l]{@{}l@{}}-the discrepancy issue can be fully \\\hspace{.5mm} resolved \end{tabular} & \begin{tabular}[l]{@{}l@{}}-long training time and a large amount\\\hspace{.5mm} of training data required for new users \end{tabular}\\\cline{2-5}
& data augmentation &\cite{soleimani2019cross088}& \begin{tabular}[l]{@{}l@{}}-can be directly applied to new \\\hspace{.5mm}users \end{tabular}& \begin{tabular}[l]{@{}l@{}} -the diversity of the synthetic data is \\\hspace{.5mm}limited and not guaranteed \end{tabular}\\\cline{2-5}
& transfer learning
& \begin{tabular}[l]{@{}l@{}} \cite{matsui2017user103}\cite{rokni2018personalized005}\\\cite{chen2019distributionally077}\cite{bai2020adversarial}\cite{chen2020metier}\end{tabular} & \begin{tabular}[l]{@{}l@{}} -less data is required for retrain \\ -common information of different \\\hspace{.5mm} users is preserved \end{tabular}& -retrain is required for each new user\\ \hline

\multirow{6}{*}{Time}   
& incremental learning & \begin{tabular}[l]{@{}l@{}}\cite{tao2016multicolumn074}\cite{gudur2019activeharnet081}\cite{saeedi2017closed105}\end{tabular} & \begin{tabular}[l]{@{}l@{}}-continuously update models to \\\hspace{.5mm}resolve the concept drift issue\end{tabular}& -few works on handling new class\\ \cline{2-5} 
& \begin{tabular}[l]{@{}l@{}}mid-level feature\\ decompose \end{tabular}& \cite{nair2019attrinet132}&\begin{tabular}[l]{@{}l@{}}-able to figure out the new class \\\hspace{.5mm}comprised with existing features\end{tabular}& \begin{tabular}[l]{@{}l@{}}-human efforts required to define mid-\\\hspace{.5mm}level features \\-unable to handle new features\end{tabular}\\ \cline{2-5}
& synthetic data & \cite{yang2019open067}&\begin{tabular}[l]{@{}l@{}}-support open-set recognition \\\hspace{.5mm}without using real out-of-set data\end{tabular}& \begin{tabular}[l]{@{}l@{}}-out-of-set data can only be recognized \\\hspace{.5mm}as one class\end{tabular}\\ \hline

\multirow{8}{*}{Sensor} 
& data augmentation &\cite{mathur2018using_023}&\begin{tabular}[l]{@{}l@{}}-can be directly applied to \\\hspace{.5mm}new sensor deployment\end{tabular}&\begin{tabular}[l]{@{}l@{}}-the diversity of the synthetic data is \\\hspace{.5mm}limited and not guaranteed\end{tabular}\\ \cline{2-5}
& how to transfer &\cite{khan2018scaling036}\cite{akbari2019transferring_024}&\begin{tabular}[l]{@{}l@{}} -less data is required for retrain \\ -common information of different \\\hspace{.5mm} users is preserved \end{tabular}& -retrain is required for each new user\\ \cline{2-5} 
& what to transfer &\cite{gjoreski2019cross133}\cite{chen2019cross089}\cite{wang2018deep090}&-select suitable source to transfer & \begin{tabular}[l]{@{}l@{}}-only feasible when multiple sources are\\\hspace{.5mm} available\end{tabular}\\ \cline{2-5} 
& \begin{tabular}[l]{@{}l@{}}domain-independent \\features\end{tabular}& \cite{zheng2019zero098}&\begin{tabular}[l]{@{}l@{}}-directly applied to new settings\end{tabular}&-only applicable to WiFi signals\\ \hline
\end{tabular}
}

\end{table}

\subsection{Composite Activity} 
Despite the success of applying a variety of deep learning models to recognizing human activities, the majority of existing research focuses on simple activities like walking, standing, and jogging, which are usually characterized by repeated actions or single body posture. The simple activities are basic and thus possess lower-level semantics. In contrast, more composite activities may contain a sequence of simple actions and have higher-level semantics, e.g., working, having dinner, and preparing coffee, which can better reflect people’s daily life. As a result, it is desirable to recognize more complicated and high-level human activities for most practical human-computer interaction scenarios. Since not only human body movements but also context information of surrounding environments are required for composite activity recognition, it is a more challenging task compared to recognizing simple activities. In addition, designing effective experiments for collecting sensor data for composite activities is also a challenging task that requires rich experience of using diverse sorts of sensors and plans of human-computer interaction applications. Therefore, the development of composite activity recognition is much more unexplored than simple activities. 

\subsubsection{Unified Models.}
Existing studies on composite activity recognition can be categorized into two streams. The first one mixes complex and simple activities and tries to create a unified model to recognize both kinds of activities. 
In \cite{vepakomma2015wristocracy}, there are twenty-two simple and composite activities attributed to four strategies: 1) Locomotive (e.g., walk indoor, run indoor); 2) Semantic (e.g., clean utensil and cooking); 3) Transitional (e.g., indoor to outdoor and walk upstairs); and 4) Postural/ relatively Stationary (e.g., standing and lying on bed). A simple multi-layer feedforward neural network was created to recognize all the activities with a high average test accuracy of 90\%. However, the results are obtained with the subject-dependent setting, where training and test samples are from the same subject, which limits the proposed method's adaptability. 

\subsubsection{ Separated Models.}
The second strategy is to consider composite activities separately from simple ones and to further regard a composite activity as the combination of a series of simple activities. This hierarchical manner is more intuitive and attracts stronger research interests. However, applying deep learning techniques to this area is still underexplored. One of the few deep learning works is \cite{peng2018aroma052} where the authors developed a multi-task learning approach to recognize both simple and composite activities simultaneously. To be concrete, the authors divided a composite activity into multiple simple activities that were represented by a series of sequential sensor signal segments. The signal segments are first input into CNNs to extract representations of low-level activities, which are then loaded into a softmax classifier for recognizing simple activities. At the same time, the CNN-extracted features of all segments are taken into an LSTM network to exploit their correlations and consequently result in a high-level semantic activity classification. In such a way, the priori of simple activities being the components of a composite activity is utilized by the shared deep feature extractor. Different from the joint learning manner, \cite{cheng2018predicting011} inferred a sequence of simple activities and its corresponding composite activity by using two conditional probabilistic models alternatively. The authors used an estimated action sequence to infer the composite activity, where the temporal correlations of simple actions are extracted for the composite activity classification. In reverse, the predicted composite activity is utilized to help derive the simple activity sequence at the next time step. As a result, the predictions of the sequence of simple activities and composite activities are mutually updated based on each other during the inference. The deep learning technique was used for feature extraction from raw signals. The experiment results showed increasing accuracy as a composite activity evolved. Even though these works have demonstrated promising solutions to recognizing composite activities, there exists a major concern that properly cutting a raw time-serial signal into segments of individual simple actions is the basis for success. A summary of the advantages and limitations of different works on composite activity recognition is presented in Table \ref{tab: ca}.

\begin{table}[ht!]
\scriptsize
\centering
\caption{Advantages and Limitations of Different Works for Composite Activity Recognition}
\begin{tabular}{|l|l|l|l|l|}
\cline{1-5}
\textbf{Treatment} & \textbf{Approaches} & \textbf{References} & \textbf{Advantages} & \textbf{Limitations} \\ \cline{1-5}
Unified & \multicolumn{1}{c|}{-} & \cite{vepakomma2015wristocracy} & -simple data collection settings & \begin{tabular}[l]{@{}l@{}} -weak generalization ability \\ -proper signal segmentation required \end{tabular}\\ \cline{1-5}
\multirow{4}{*}{Separated} & joint learning & \cite{peng2018aroma052} & \begin{tabular}[l]{@{}l@{}} -simultaneously recognizing simple \\ \hspace{.5mm}and composite activity \\ -mutual performance enhancement \end{tabular} & \begin{tabular}[l]{@{}l@{}} -priori knowledge required \\ -poor adaptability \end{tabular} \\ \cline{2-5}
& action to activity & \cite{cheng2018predicting011} & \begin{tabular}[l]{@{}l@{}} -intuitive \\ -favorable adaptability \\ -mutual performance enhancement\end{tabular} & \begin{tabular}[l]{@{}l@{}} -complex training scheme and inference \\ \hspace{.5mm} process  \end{tabular}\\ \cline{1-5}
\end{tabular}
\label{tab: ca}
\end{table}
\vspace{-0.5cm}
\subsection{Data Segmentation}
As original sensor data is represented by continuously streaming signals, a fixed-size window is always used to partition raw sensor data sequences into segments as input into a model for activity recognition. This is essential to overcome the limitation of the sample of a single time step to provide adequate information about an activity. Ideally, one partitioned data segment 
processes only one activity, and thus a model predicts a single label for all the samples within a single window. However, the samples in one window may not always share the same label when an activity transition occurs in the middle of the window. Therefore, an optimal segmentation approach is critical to increasing activity recognition accuracy.
\subsubsection{Explicit Segmentation.}
An intuitive manner is to attempt various fixed window sizes empirically. Nevertheless, although a larger window size provides richer information, it increases the possibility that a transition occurs in the middle of windows. On the contrary, a smaller window size cannot afford enough information. In light of this issue, \cite{akbari2018hierarchical040} reported a hierarchical signal segmentation method, which initially used a large window size and gradually narrowed down the segmentation until only one activity is in a sub-window. The narrow-down criterion is that two consecutive windows have different labels or the classification confidence is less than a threshold. Different from the hierarchical framework, some researchers explored to directly assign a label for each time-step instead of predicting a window as a whole \cite{yao2018efficient066,zhang2018human092}. Inspired by semantic segmentation in the computer vision community, the authors employed fully connected networks (FCNs)\cite{long2015fully} to achieve such a goal. Data from a large window size is input, and a 1D CNN layer is used to replace the final softmax layer, where the length of the feature map equals to time steps and the number of the feature maps equals to the number of activity classes, to predict a label for each time step. Therefore, the FCNs could not only use the information of the corresponding time step itself but also utilize the information of its neighboring time steps. 

\subsubsection{Implicit Segmentation.}
Explicit segmentation for activity recognition is not practical since users performing activities in unfixed durations. 
In \cite{varamin2018deep084}, Varamin et al. defined unsegmented activity recognition as a set prediction problem. They designed a multi-label architecture to simultaneously predict the number of ongoing activities and the occurring possibility of each alternative activity without explicit segmentation. 
Table \ref{tab: ds} summarizes the advantages and limitations of different methods for data segmentation.

\begin{table}[ht!]
\scriptsize
\centering
\caption{Advantages and Limitations of Different Works for Data Segmentation}
\scalebox{0.97}{
\begin{tabular}{|l|l|l|l|l|}
\cline{1-5}
\textbf{Treatment} & \textbf{Approaches} & \textbf{References} & \textbf{Advantages} & \textbf{Limitations} \\ \cline{1-5}
\multirow{5}{*}{Explicit segmentation} & \begin{tabular}[l]{@{}l@{}}hierarchical \\ narrow-down \end{tabular} & \cite{akbari2018hierarchical040} & \begin{tabular}[l]{@{}l@{}} -able to deal with a transition within \\ \hspace{.5mm}a window \\ -able to capture long range information\end{tabular} & \begin{tabular}[l]{@{}l@{}} -limited generalization ability\\ -multiple classifiers required \\ -limited in capturing transitions\end{tabular} \\ \cline{2-5}
& time-step wise & \cite{yao2018efficient066}\cite{zhang2018human092} & \begin{tabular}[l]{@{}l@{}} -able to deal with a transition within\\ \hspace{.5mm}a window \\ -able to capture long range information\\-fine grained segmentation\end{tabular} & \begin{tabular}[l]{@{}l@{}} -difficult to define exact transition\\\hspace{.5mm} periods for ground truth\end{tabular}\\ \cline{1-5}

Implicit segmentaion & \begin{tabular}[c]{@{}c@{}} multi-label \end{tabular} & \cite{varamin2018deep084} & \begin{tabular}[l]{@{}l@{}}-simple structure and training scheme \\ -able to capture long range information \end{tabular}& \begin{tabular}[l]{@{}l@{}} -relatively coarse \\ -not able to capture transitions \\ -not able to identify activity\\ \hspace{.5mm} sequence within a window \end{tabular}\\ \cline{1-5}
\end{tabular}
}
\label{tab: ds}
\end{table}

\subsection{Concurrent Activity}
In real-world scenarios, in addition to performing each activity one after another in a sequential fashion, a person may carry out more than one activity at the same time, which is called concurrent activities. For instance, one may make a phone call when watching TV. From the angle of sensor signals, a piece of data may correspond to multiple ground truth labels. Therefore, concurrent activity recognition can be abstracted as a multi-label task. Note that the concurrent activity is executed by a single subject.

\subsubsection{Recognize Individually.}
A concurrent activity can be considered as several individual activities.
Zhang et al. \cite{zhang2017car_021} designed an individual fully-connected network for each candidate activity on top of shared multimodal fusion features. The final decision-make layer classified each activity independently by independent softmax layers. A key drawback of this kind of structure is that the computational cost would increase considerably with the number of activities rises. To resolve this issue, the authors further proposed to use a single neuron with the $sigmoid$ activation to make binary classification (performed or not) for each activity \cite{li2017concurrent101}. 

\subsubsection{Recognize Concurrently.}
In contrast, Okita and Inoue \cite{okita2017recognition027} also targeted the concurrent activities, but directly considering the possibility of different activities occurring concurrently. They suggested a multi-layer LSTM framework to give the concurrent possibility of every possible activity combination. The main limitation of this work is the output dimension would explode exponentially as the increase of the amount of concurrent activities. The pace of exploring deep learning methods on concurrent activity recognition is still slow, and there is a large room to improve.  A summary of the advantages and limitations of different approaches for concurrent activity recognition is illustrated in Table \ref{tab: cona}.

\begin{table}[ht!]
\scriptsize
\centering
\caption{Advantages and Limitations of Different Works for Concurrent Activity Recognition}
\begin{tabular}{|l|l|l|l|l|}
\cline{1-5}
\textbf{Treatment} & \textbf{Approaches} & \textbf{References} & \textbf{Advantages} & \textbf{Limitations} \\ \cline{1-5}
 Individually & \begin{tabular}[l]{@{}l@{}}multi-label \end{tabular} & \cite{li2017concurrent101}\cite{zhang2017car_021} & \begin{tabular}[l]{@{}l@{}} -simple architecture\end{tabular} & \begin{tabular}[l]{@{}l@{}} -limited adaptability to new activities \end{tabular} \\ \cline{1-5}

 Concurrently & \begin{tabular}[l]{@{}l@{}} multi-layer LSTM and \\ high dimensional tensor \end{tabular} & \cite{okita2017recognition027} & \begin{tabular}[l]{@{}l@{}}-achieve results directly \end{tabular}& \begin{tabular}[l]{@{}l@{}} -computational cost increases exponentially \\ \hspace{1mm}with the number of activities increases \\ -limited adaptability to new activities \end{tabular}\\ \cline{1-5}
\end{tabular}
\label{tab: cona}
\end{table}

\subsection{Multi-occupant Activity}
Most of the state-of-the-art works focus on monitoring and assisting people with regard to single-occupant. Nevertheless, living and working spaces are usually resided by multiple subjects; hence, designing solutions for handling multi-occupant is of notably practical significance. There are mainly two types of multi-occupant activities: \textit{parallel activity} where occupants perform activities individually such as one occupant is eating while the other one is watching TV and \textit{collaborative activity} where multiple occupants collaborate together to perform the same activity such as two subjects play table tennis \cite{benmansour2015multioccupant}. For the \textit{parallel activity} recognition, when only wearable-sensors are used, it can be divided into multiple single-occupant activity recognition tasks and solved by conventional solutions; while ambient or object sensors are used, data association of mapping sensed signals to the occupant who actually causes the generation of the data becomes the major challenge, which gets more serious as the number of occupants in the space increases. The problem of data association is crucial to the multi-occupant scenario since failing to do so, data would be useless and could even endanger the life of residents in telehealth applications. For the \textit{collaborative activity}, human interactions and instruments are generally involved; thus, context and object-use information play vital roles in designing recognition solutions. Although the multi-occupant activity recognition is of great meaning, its deep learning-based research is still limited.

\subsubsection{Collaborative Activity.}
In \cite{rossi2018multimodal063}, both wearable and ambient sensors were used to recognize group activities of two occupants. The ambient sensors were leveraged for extracting context information which is represented by disparate functional indoor areas. The sensor data of different occupants was input into different RBMs separately and then merged into a sequential network, a DBN and an MLP, for the inference of the group activity. Pretty high accuracy of nearly 100\% was achieved. However, most of their targeting scenarios are constrained with two occupants performing the same activity together. 

\subsubsection{Parallel Activity.}
On the contrary, Tran et al. \cite{tran2018multi035} did not restrain the occupants to act together. They aimed at recognizing activities for each occupant separately. A multi-label RNN was created with each RNN cell responding to activity recognition of one occupant. Nevertheless, the authors only used ambient sensors and did not propose a specific solution to the data association issue. 
Table \ref{tab: ma} summarizes the advantages and limitations of different methods for multi-occupant activity recognition.

\begin{table}[ht!]
\scriptsize
\centering
\caption{Advantages and Limitations of Different Works for Multi-occupant Activity Recognition}
\begin{tabular}{|l|l|l|l|l|}
\cline{1-5}
\textbf{Targeting scenario} & \textbf{Sensors} & \textbf{References} & \textbf{Advantages} & \textbf{Limitations} \\ \cline{1-5}
Collaborative activity& \begin{tabular}[l]{@{}l@{}}ambient and wearable \end{tabular} & \cite{rossi2018multimodal063} & \begin{tabular}[l]{@{}l@{}} -nearly 100\% recognition \\ \hspace{.5mm} accuracy\end{tabular} & \begin{tabular}[l]{@{}l@{}} -occupants are constrained to \\ \hspace{.5mm} perform the same activity together\end{tabular} \\ \cline{1-5}

Parallel activity & \begin{tabular}[l]{@{}l@{}} ambient \end{tabular} & \cite{tran2018multi035} & \begin{tabular}[l]{@{}l@{}}-no constraints to occupants \end{tabular}& \begin{tabular}[l]{@{}l@{}} -unable to associate activities to \\ \hspace{.5mm} occupants \end{tabular}\\ \cline{1-5}
\end{tabular}
\label{tab: ma}
\end{table}

\subsection{Computation Cost}
Although deep learning models have shown dominant accuracy in the sensor-based human activity recognition community, they are typically resource-intensive. For example, the early DCNN architecture, AlexNet \cite{krizhevsky2012imagenet}, which has five CNN layers and three fully-connected layers, processes 61M parameters (249MB of memory) and performs 1.5B high precision operations to make a prediction. For non-portable applications, Graphic Processing Units (GPUs) are usually leveraged to accelerate computation. However, GPUs are very expensive and power-hungry so that not suitable for real-time applications on mobile devices. Moreover, current research has demonstrated that making a neural network deeper by introducing additional layers and nodes is a critical approach to improving model performance, which inevitably increases computational complexity. Therefore, it is essential and challenging to resolve the issue of high computation cost to realize real-time and reliable human activity recognition on mobile devices by deep learning models. 

\subsubsection{Layer Reduction.}
Considering deep neural networks are more effective in feature extraction than shallow ones, a combination of human-crafted and deep features is a potential solution to lowering computation cost. In \cite{ravi2016deep107}, the authors incorporated the spectrogram features with only one CNN layer and two fully-connected layers for human activity recognition. The hybrid architecture showed comparative recognition accuracy to state-of-the-art methods through evaluation on four benchmark datasets. To validate the feasibility of real-time usage, the authors implemented the proposed method on three different mobile platforms, including two smartphones and one on-node unit. The results revealed milliseconds to tens of milliseconds computational time of one prediction suggesting the possibility of real-time applications. \cite{pires2018multi012} also demonstrates the combination of hand-crafted features and a neural network is a potential plan to achieve real-time activity recognition on a mobile device. In addition to the cascade structure of hand-crafted features and deep learning features, \cite{ravi2016deep113} proposed to arranged the deep learning features and hand-crafted features in parallel before fed into a fully-connected classifier. This structure could increase recognition accuracy with only a small gain of computational consumption. 

\subsubsection{Network Optimization.}
Optimizing basic neural network cells and structure is another intuitive scheme of decreasing computation complexity. In \cite{vu2017self116}, Vu et al. used a self-gated recurrent neural network (SGRNN) cell to decline the complexity of a standard LSTM and prevent gradient vanishing. Their experiments displayed superior computation efficiency to LSTM and GRU in terms of the running time and model size. However, the running time was still in the order of hundreds of milliseconds and no real-world evaluation on mobile devices is carried out to show possible real-time implementation. For CNN-based methods, reducing filter size is an effective means to optimize the memory consumption and the number of computation operations. For example, \cite{ravi2016deep113} utilized 1D-CNNs instead of 2D-CNNs to control the model size. 
A more insightful strategy to dealing with both the storage and computational problems is the quantization of network \cite{edel2016binarized109}. This scheme is to constraint the weights and outputs of activation functions to two discrete values (e.g., -1, +1) instead of continuous numbers. There are three major benefits of network quantization: 1) the memory usage and model size are greatly reduced when compared to the full and precise networks; 2) the bitwise operations are considerably more efficient than conventional floating or fixed-point arithmetic; 3) if bitwise operations are used, most multiply-accumulate operations (require hundreds of logic gates at least) can be replaced by popcount-XNOR operations (only require a single logic gate), which are especially well suited for FPGAs and ASICs \cite{yang2018dfternet068}. In \cite{yang2018dfternet068}, Yang et al. explored a 2-bit CNN with weights and activation constrained to \{-0.5, 0, 0.5\} for efficient activity recognition.
Table \ref{tab: cc} summarizes the advantages and limitations of different methods for reducing computation cost.

\begin{table}[ht!]
\scriptsize
\centering
\caption{Advantages and Limitations of Different Works for Computation Cost}
\scalebox{0.97}{
\begin{tabular}{|l|l|l|l|l|}
\cline{1-5}
\textbf{Solution scheme} & \textbf{Approaches} & \textbf{References} & \textbf{Advantages} & \textbf{Limitations} \\ \cline{1-5}
Layer reduction & \begin{tabular}[l]{@{}l@{}}combination of hand-crafted \\ features and deep features\end{tabular} & \cite{pires2018multi012}\cite{ravi2016deep113}\cite{ravi2016deep107} & \begin{tabular}[l]{@{}l@{}} -simple structure\\-incorporate features of \\ \hspace{.3mm} different aspects \end{tabular} & \begin{tabular}[l]{@{}l@{}} -domain knowledge required for\\ \hspace{.5mm}hand-crafted features \\-complex preprocessing \end{tabular} \\ \cline{1-5}

\multirow{5}{*}{Network optimization} & \begin{tabular}[l]{@{}l@{}} optimizing basic block \end{tabular} & \cite{vu2017self116}\cite{ravi2016deep113} & \begin{tabular}[l]{@{}l@{}}-end-to-end manner\end{tabular}& \begin{tabular}[l]{@{}l@{}} -limited computation cost\\ \hspace{.5mm}  reducing capability \end{tabular}\\ \cline{2-5}

& network quantization & \cite{yang2018dfternet068}\cite{edel2016binarized109} & \begin{tabular}[l]{@{}l@{}}-powerful computation \\ \hspace{.5mm}cost reducing capability \\ -suitable for FPGAs and \\ \hspace{.5mm}ASICs\end{tabular}& \begin{tabular}[l]{@{}l@{}} -risk of performance  degradation\end{tabular}\\ \cline{1-5}
\end{tabular}
}
\label{tab: cc}
\end{table}

\subsection{Privacy}
The main application of human activity recognition is to monitor human behaviors so the sensors capture the activities of a user continuously. Since the way an activity is performed varies among users, it is possible for an adversary to infer user sensitive information such as age through the time series sensor data. Specifically, for the deep learning technique, its black-box characteristic may be at the risk of revealing user-discriminative features unintentionally. In \cite{iwasawa2017privacy014}, the authors investigated the privacy issue of using CNN features for human activity recognition. Their empirical studies revealed that although CNN is trained with a cross-entropy loss only targeting activity classification, the obtained CNN features still showed powerful user-discriminative ability. A simple logistic regressor could achieve a high user-classification accuracy of 84.7\% when using the CNN features basically extracted for activity while the same classifier could only obtain 35.2\% user-classification accuracy on raw sensor data. Therefore, it is essential to address the privacy leakage potentials of a deep learning model originally used for human activity recognition.

\subsubsection{Transformation.}
To address this concern, some researchers explored to utilize an adversarial loss function to minimize the discriminative accuracy of specific privacy information during the training process. For example, Iwasawa et al. \cite{iwasawa2017privacy014} proposed to integrate an adversarial loss with the standard activity classification loss to minimize the user identification accuracy. The authors of \cite{malekzadeh2019mobile} and \cite{malekzadeh2018protecting} also adopted the similar idea to prevent privacy leakage. Their experiment results show an effective reduction of inferring accuracy for sensitive information. However, an adversarial loss function can only be used for protecting one kind of private information, such as user identity and gender. In addition, the adversarial loss goes against the end-to-end training process that making it hard to converge stably. Considering this gap, \cite{zhang2019icdm} borrowed the idea of image style transformation from the computer vision community to protect all private information at once. The authors creatively viewed raw sensor signals from two aspects: "style" aspect that describes how a user performs an activity and was influenced by user's identical information like age, weight, gender, height, et al.; "content" aspect that describes what activity a user performs. They proposed to transform raw sensor data to have the "content" unchanged but the "style" is similar to random noises. Therefore, the method has the potential to protect all sensitive information at once. 

\subsubsection{Perturbation.}
Besides the data transformation strategy, data perturbation is another way to resolve the privacy issue. For example, Lyu et al. proposed to tailor two kinds of data perturbation mechanisms: Random Projection and repeated Gompertz to achieve a better tradeoff between privacy and recognition accuracy \cite{lyu2017privacy056}. Recently, differential privacy has gained increasing research attention due to its strong theoretical privacy guarantee. Phan et al. \cite{phan2016differential008} proposed to perturb the objective functions of the traditional deep auto-encoder to enforce the $\epsilon$-differential privacy. In addition to the privacy preservation in feature extraction layers, an $\epsilon$-differential privacy preserving softmax layer was also developed for either classification or prediction. Different from the above approaches, this method provided theoretical privacy guarantees and error bounds. The advantages and limitations of different methods for protecting user privacy in activity recognition are in Table \ref{tab: pp}.

\begin{table}[ht!]
\scriptsize
\centering
\caption{Advantages and Limitations of Different Works for Privacy Protection}
\scalebox{0.98}{
\begin{tabular}{|l|l|l|l|l|}
\cline{1-5}
\textbf{Protection scheme} & \textbf{Approaches} & \textbf{References} & \textbf{Advantages} & \textbf{Limitations} \\ \cline{1-5}

\multirow{5}{*}{Transformation} & \begin{tabular}[l]{@{}l@{}}adversarial training \end{tabular} & \cite{iwasawa2017privacy014}\cite{malekzadeh2018protecting}\cite{malekzadeh2019mobile} & \begin{tabular}[l]{@{}l@{}} -simple network structure \end{tabular} & \begin{tabular}[l]{@{}l@{}} -unstable training \\ -sensitive labels required \\ -new structure needed for new \\\hspace{.5mm} privacy information \end{tabular} \\ \cline{2-5}

& \begin{tabular}[l]{@{}l@{}} style transfer \end{tabular} & \cite{zhang2019icdm} & \begin{tabular}[l]{@{}l@{}} -protect all privacy information \\ \hspace{.5mm} at one transformation \\ -free of sensitive information for \\\hspace{.5mm} training \end{tabular} & \begin{tabular}[l]{@{}l@{}} -complex structure and training \\\hspace{.5mm}strategy \end{tabular} \\ \cline{1-5}

\multirow{3}{*}{Perturbation} & \begin{tabular}[l]{@{}l@{}} direct noise insertion \end{tabular} & \cite{lyu2017privacy056} & \begin{tabular}[l]{@{}l@{}}-simple\end{tabular}& \begin{tabular}[l]{@{}l@{}} -limited ability to retain activity \\\hspace{.5mm}information \end{tabular}\\ \cline{2-5}

& differential privacy & \cite{phan2016differential008} & \begin{tabular}[l]{@{}l@{}}-theoretical privacy guarantees \\\hspace{.5mm}and error bounds \end{tabular}& \begin{tabular}[l]{@{}l@{}} -only validated on fully connected \\\hspace{.5mm}layers\end{tabular}\\ \cline{1-5}
\end{tabular}
}
\label{tab: pp}
\end{table}

\subsection{Interpretability}

Sensory data for human activity is unreadable. A data sample may include diverse modalities (e.g., acceleration, angular velocity) from multiple positions (e.g., wrist, ankle) in a time window. However, only a few of modalities from specific positions contribute to identifying certain activities \cite{kwon2015analysis099}. For example, lying is distinguishable when people are horizontal (magnetism), and ascending stairs can be recognized by the forward and the upward acceleration of people's ankle. Unrelated modalities can introduce noise and deteriorate the recognition performance. Moreover, the significance of each modality changes over time. For instance, in a Parkinson disease detection system, anomaly only appears in gait in a short period instead of the entire time window \cite{zeng2018understanding030}. Intuitively, the modality shows more considerable significance when the corresponding body part is actively moving.

Despite the success of deep learning in activity recognition, the inner mechanisms of deep learning networks still remain unrevealed. Considering the varying salience of modalities and time intervals, it is necessary to interpret the neural networks to explore the factors of the models' decisions. For example, when a deep learning model identifies that a user is walking, we tend to know which modality from which time interval is the determinant. Therefore, the interpretability of deep learning methods has become a new trend in the human activity recognition community.

\subsubsection{Feature Visualization.}
The basic idea of interpretable deep learning is to automatically decide the importance of each part of the input data, and to achieve high accuracy by omitting the unimportant parts and focusing on the salient parts. In fact, the standard fully connected layers already possess such capacity as they automatically reduce the weights of less important neurons during training, but we still need to visualize the features for interpretation.
Some researchers \cite{xue2018understanding095,brophy2018interpretable085} visualized the features extracted by neural networks. Salient features are sent to the subsequent models after the authors find out their relationships to the activities from the visualization \cite{xue2018understanding095}. Nutter et al. \cite{nutter2018design062} transformed sensory data to images so that visualization tools can be applied to the sensory data for more direct interpretability.

\subsubsection{Attentive Selection.}
\textit{Attention mechanism} is recently popular in deep learning areas and is originally a concept in biology and psychology that illustrates how we restrict our attention to something crucial for better cognitive results. Inspired by this, 
researchers apply neural attention mechanisms to deep learning to give neural networks the capability of concentrating on a subset of inputs that really matters.
Since the principle of deep attention models is to weigh input components, components with higher weights are assumed to be more tightly related to the recognition task and show greater influence over the models' decisions \cite{serrano2019attention}. Some works employed attention mechanism to interpret deep model behaviors \cite{zhang2019graph,zhang2018ready,zhang2020motor}. 
Back to human activity recognition, attention mechanism not only highlights the most distinguishable modalities and time intervals but also informs us of the most contributing modalities and body parts to specific activities.
Deep attention approaches can be categorized into soft attention and hard attention based on their differentiability.

\textit{\textbf{Soft Attention.}}
In machine learning, ``soft'' means differentiable. Soft attention assigns weight from $0$ to $1$ to each element of the inputs. It decides how much attention to focus on each element. Soft attention uses softmax functions in the attention layers to compute the weights so the whole model is fully differentiable where gradients can be propagated to other parts of the network \cite{zhang2019convolutional}. Attention layers can be inserted into sequence-to-sequence LSTMs for feature extraction \cite{tang2016sequence019}.
Attention layers can also be inserted in the neural networks to tune the weights of all samples\cite{murahari2018attention029} in sliding windows since samples at different time points have varying contributions to activity recognition. Shen et al. \cite{shen2018sam093} further considered the temporal context. They designed a segment-level attention approach to decide which time segment contains more information. Combined with gated CNN, the segment-level attention better extracts temporal dependencies. Zeng et al. \cite{zeng2018understanding030} developed attention mechanisms in two perspectives. They first propose sensor attention on the inputs to extract the salient sensory modalities and then apply temporal attention to an LSTM to filter out the inactive data segments. Spatial and temporal attention mechanisms are employed in \cite{ma2019attnsense127}. Especially, the spatial dependencies are extracted by fusing the modalities with self-attention.

\textit{\textbf{Hard Attention.}}
Hard attention determines whether to attend to a part of inputs or not. The weight assigned to an input part is either $0$ or $1$ so the problem is non-differentiable.
The process involves making a sequence of selections about which part to attend. 
The selection can be output by a neural network. However, since there is no ground truth indicating the correct selection policy, hard attention should be represented as a stochastic process. This is where \textit{deep reinforcement learning} comes in. Deep reinforcement learning tackles the selection problems in deep learning and allows the models to propagate gradients in the space of selection policies.

Different reinforcement learning techniques can be applied to hard attention mechanisms in human activity recognition. Zhang et al. \cite{zhang2018multi013} use dueling deep Q networks as a core of hard attention to focus on the salient parts of multimodal sensory data. Chen et al. \cite{chen2018interpretable060,chen2019semisupervised125} mined important modalities and elide undesirable features with policy gradient. The attention is embedded into an LSTM to make selections step by step because LSTM incrementally learns information in an episode. Chen et al. \cite{chen2019multi078} further considered the intrinsic relations between activities and sub-motions from human body parts. They employ multiple agents to concentrate on modalities that are related to sub-motions. Multiple agents coordinate to portray the activities. The visualization of the selected modalities and body parts validates that the attention mechanism provides insights into how sensory data elements affect the models' prediction of activities.
 The advantages and limitations of different methods for model interpretability are listed in Table \ref{tab: i} .

\begin{table}[ht!]
\scriptsize
\centering
\caption{Advantages and Limitations of Different Works for Model Interpretability}
\scalebox{0.98}{
\begin{tabular}{|l|l|l|l|l|}
\cline{1-5}
\textbf{Interpretation scheme} & \textbf{Approaches} & \textbf{References} & \textbf{Advantages} & \textbf{Limitations} \\ \cline{1-5}
Feature visualization & \multicolumn{1}{c|}{-} & \cite{brophy2018interpretable085}\cite{nutter2018design062}\cite{xue2018understanding095}& \begin{tabular}[l]{@{}l@{}} -adopt current tools of \\\hspace{.5mm}computer vision \\ -simple and intuitive\end{tabular}& \begin{tabular}[l]{@{}l@{}} -unable to interpret hidden layers \\ -limited power compared to visualize \\\hspace{.5mm} images as raw signals are unreadable\end{tabular} \\ \cline{1-5}

\multirow{4}{*}{Attentive selection} & \begin{tabular}[l]{@{}l@{}} soft attention \end{tabular} & \cite{ma2019attnsense127}\cite{malekzadeh2018protecting}\cite{murahari2018attention029}\cite{zeng2018understanding030} & \begin{tabular}[l]{@{}l@{}}-fully differentiable \\-applied to both temporal\\\hspace{.4mm} and modality selection \\\hspace{.9mm}interpretation\end{tabular}& \begin{tabular}[l]{@{}l@{}} -high cost when input is large \end{tabular}\\ \cline{2-5}

& hard attention & \cite{chen2018interpretable060}\cite{chen2019semisupervised125}\cite{chen2019multi078}\cite{zhang2018multi013}& \begin{tabular}[l]{@{}l@{}}-less calculation during test \end{tabular}& \begin{tabular}[l]{@{}l@{}} -complex training procedure \\-applied only to modality selection \\\hspace{.5mm} interpretation\end{tabular}\\ \cline{1-5}
\end{tabular}
}
\label{tab: i}
\end{table}

\section{Future Research Direction}

To develop full potential of deep learning in human activity recognition, some future research directions are worthy of further investigation. Future directions can be stimulated by the challenges summarized in this work. Despite the effort devoted to these challenges, some of them are still not fully explored such as class imbalance, composite activities, concurrent activities, etc. Although current research works still lack comprehensive and reliable solutions for the challenges, they lay concrete foundations and show guidance for future directions. 

Moreover, there are other research directions that have rarely been explored before. We outline several key research directions that urgently need to be exploited as follows.

\begin{itemize}[leftmargin=*]
    \item \textbf{\textit{Independent unsupervised methods.}} Human activity recognition needs a sufficient amount of annotated samples to train the deep learning models. Unsupervised learning can help mitigate such requirements. So far, deep unsupervised models used for human activity recognition are mainly used for extracting features but are not able to identify activities because there is no ground truth. Therefore, one potential method for unsupervised learning to infer true labels is to seek other knowledge, which leads us to a popular method, \textit{deep unsupervised transfer learning} \cite{bengio2012deep}. Another way is to resort to \textit{data-driven methods} such as ontology \cite{riboni2011ontology}.

    \item \textbf{\textit{Identifying new activities.}} Identifying novel activities that have never been seen by the models is a big challenge in human activity recognition. A reliable model should be able to learn the new knowledge online and achieve accurate recognition without any ground truth. A promising way is to learn features that are scalable to diverse activities. While \cite{nair2019attrinet132} enlightens us that \textit{mid-level attributes} can be used to depict activities with a set of characteristics, \textit{disentangled features} \cite{tran2017disentangled} may be another serviceable solution to representing novel activities.
    
    \item \textbf{\textit{Future activity prediction.}} Future activity prediction is an extension of activity recognition. Unlike activity recognition, the activity prediction system can forecast users' behaviors in advance. The prediction system is useful in detecting human intention so it can be applied to smart services, criminal detection and driver behavior prediction.
    In some common behavior tasks, the activities are usually in a certain order. Therefore, modeling the temporal dependencies across activities is beneficial to predict future predictions. \textit{LSTMs} \cite{baccouche2010action} are suitable for such tasks. But for long-span activities, LSTMs cannot contain such long dependencies. In this case, \textit{intention recognition} based on brain signals \cite{zhang2018cascade} can assist to inspire activity prediction.
    
    \item \textbf{\textit{A standardization of the state-of-the-art.}} While hundreds of works have been investigated in deep learning and sensor-based human activity recognition, there lacks a standardization of the state-of-the-art for a fair comparison. The experiment settings and evaluation metrics for assessing the performance of activity recognition vary from paper to paper. While deep learning heavily relies on the training data, the division of training/ test/ validation sets influences the recognition results. Other factors including data processing and the implementation platforms also lead to skewed comparison. Therefore, having a mature standardization for all researchers is pressing.
    It is noteworthy that such an issue is absent in other areas. For example, ImageNet Challenge \cite{russakovsky2015imagenet} meticulously defines details in the experiment setting to ensure impartial comparison. 
    Jordao et al. \cite{jordao2018human096} implemented and evaluated a set of existing works with standardized settings, but there is still no rigorous and well-recognized standardization in the field of human activity recognition.

\end{itemize}

\section{Conclusion}

This work aims at suggesting a rough guideline for novices and experienced researchers who have interest in deep learning methods for sensor-based human activity recognition.
We present a comprehensive survey to summarize the current deep learning methods for sensor-based human activity recognition. 
We first introduce the multi-modality of the sensory data and available public datasets and their extensive utilization in different challenges. 
We then summarize the challenges in human activity recognition based on their reasons and analyze how existing deep methods are adopted to address the challenges. At the end of this work, we discuss the open issues and provide some insights for future directions.

\bibliographystyle{ACM-Reference-Format}
\bibliography{survey}

\end{document}